\documentclass[10.9pt,twocolumn]{article}
\pdfoutput=1

\usepackage{slashed}
\usepackage{xcolor, verbatim}
\usepackage{latexsym}
\usepackage{amsmath}
\usepackage{amssymb}
\usepackage{widetext}
\usepackage{cite}
\usepackage{graphicx}
\graphicspath{{figs/}}
\usepackage{bm}
\usepackage{adjustbox}
\usepackage{multirow}
\usepackage{rotating}

\newcommand{\vtext}[1]{\begin{sideways}\small{#1}\end{sideways}}

\makeatletter
\g@addto@macro\bfseries{\boldmath}
\makeatother

\begin{document}

\title{\textsc{Effective field theory for vector-like leptons\\ and its 
collider signals}\\[0.2cm]}
\author{
\textbf{Mikael Chala$^{\,a}$, Pawe\l{} Koz\'ow$^{\,a}$, Maria Ramos$^{\,a,b}$ 
and Arsenii 
Titov$^{\,c}$}\\[0.6cm]
$^a$\textit{CAFPE and Departamento de F\'isica Te\'orica y del Cosmos,}\\
\textit{Universidad de Granada, E--18071 Granada, Spain}\\[0.2cm]
$^b$\textit{Laborat\'orio de Instrumenta\c{c}\~ao e F\'isica Experimental 
de Part\'iculas,}\\
\textit{Departamento de F\'isica da Universidade do Minho,}\\
\textit{Campus de Gualtar, 4710-057 Braga, Portugal}\\[0.2cm]
$^c$\textit{Dipartimento di Fisica e Astronomia ``G. Galilei'',}\\
\textit{Universit\`a degli Studi di Padova and INFN, Sezione di Padova,} \\
\textit{Via Francesco Marzolo 8, I--35131 Padova, Italy}
}

\date{}

\twocolumn[
\begin{@twocolumnfalse}
\maketitle
\begin{abstract}
We argue that in models with several high scales; 
\textit{e.g.} in composite Higgs models or in gauge extensions of the 
Standard Model (SM), vector-like leptons can be likely produced in 
a relatively 
large $\sqrt{s}$ region of the phase space. Likewise, they can easily decay 
into 
final states not containing SM gauge bosons. This contrasts with 
the 
topology in which these new particles are being searched for at the LHC. 
Adopting an effective field theory approach, we show 
that searches for excited leptons must be used instead to test this scenario. 
We derive bounds on  all the relevant interactions of dimension six; the most 
constrained ones being of 
about $0.05$ TeV$^{-2}$. We build new observables to improve current analyses 
and study 
the impact on all single-field UV completions of the SM 
extended with a vector-like lepton that can 
be captured by the effective field theory at tree level, in the current and in 
the high-luminosity phase of the LHC.\\
\end{abstract}
\end{@twocolumnfalse}
]

\section{Introduction}
%
Leptons beyond those of the Standard Model (SM), if they exist, have masses 
well above the electroweak (EW) scale, or else they would conflict
with EW and Higgs precision data. Therefore, they can not get their masses from the Higgs mechanism. Instead, any such new lepton $E$ must be
vector-like with respect to the SM gauge symmetry 
$SU(3)_c\times SU(2)_L\times U(1)_Y$; namely
the left-handed (LH) and right-handed (RH) components transform in the same 
representation, which allows an explicit mass term $M_E \overline{E}E$.

Direct searches for vector-like leptons (VLLs) are being performed at the 
LHC~\cite{Aad:2015cxa,Khachatryan:2015scf,Aad:2015dha,
Grancagnolo:2015zsh,Sirunyan:2019ofn,Aad:2020fzq,Sirunyan:2019bgz}, with null results so far. 
This observation does not necessarily imply that 
there are no VLLs below the TeV scale.
It can rather be that, contrary to what all the aforementioned experimental
analyses presume, the actual VLLs \textit{(i)} are mostly single produced, 
\textit{(ii)} populate mainly
the phase space of relatively large
$\sqrt{s}$ and \textit{(iii)} do not decay to SM gauge bosons.

This is indeed the case in several theoretical frameworks; 
for example in some composite Higgs models (CHMs). 
The latter involve a new strong sector that
confines around the scale $f_*\sim$ TeV. While
vector resonances are expected to have masses of order $\Lambda\sim g_* f_*$, with $g_*\gg 1$ being the coupling between composite resonances, 
fermionic resonances should rather lie at a scale closer to $f_*$, 
generating the hierarchy $m_E\ll \Lambda$. 
One reason is that EW precision data (EWPD) and flavour
constraints are much stronger for vector than for fermionic 
resonances~\cite{Chala:2014mma}. 
One additional reason is that the Higgs mass in CHMs is much
more sensitive to $m_E$ than to $m_V$, particularly in those in which the 
SM leptons interact sizeably with the strong sector. 
Such models, in turn, are motivated by the flavour 
anomalies~\cite{Niehoff:2015bfa,Niehoff:2015iaa,Carmona:2015ena,Carmona:2017fsn,Sannino:2017utc,Chala:2018igk}.

It is therefore likely that the actual phenomenology of VLLs at the LHC 
must be described by an effective field theory (EFT)~\footnote{For  
recent studies of the impact of higher-dimensional 
operators on the phenomenology of vector-like quarks, see 
Refs.~\cite{Criado:2019mvu,Kim:2018mks,Alhazmi:2018whk}.}, including not
only the dimension-four Yukawa interaction $\sim y \overline{l_L} H E$ (with 
$l_L$ and $H$ being 
the LH lepton doublet and the Higgs boson,
respectively), but also dimension-six interactions suppressed by $1/\Lambda^2$. We note however that $y$ modifies the $Z$ coupling to the SM
leptons and it is therefore very constrained by EWPD; 
$y\lesssim 0.1$~\cite{deBlas:2013gla}. Moreover, in CHMs, $y\sim Y_* \rm s_L 
s_R$ where $Y_*$ is a proto-Yukawa coupling between the fully composite Higgs 
and the fermionic resonances and $\rm s_L$ and $\rm s_R$ are the degree of 
compositeness 
of LH and RH fermions. If Minimal Flavour Violation is at work, then $s_L$ 
vanishes~\cite{Redi:2013pga}, and therefore $y\to 0$.
Hence,
the single production $pp\to E\ell$ mode populating the
relatively large $\sqrt{s}$ phase space dominates, 
because the production cross section grows as $\sigma\sim s/\Lambda^4$.

Likewise, the non-resonant decay  channel $E\to \ell q\overline{q}$ 
can dominate over $E\to Z/h\ell$ (or $W\nu$). 
Naive dimensional analysis tells us that this happens 
provided $y\lesssim 0.1 (m_E/\Lambda)^2$.~\footnote{Note also that 
effective Higgs operators, \textit{e.g.} $(H^\dagger iD_\mu 
H)(\overline{e}\gamma^\mu e)$ or $(H^\dagger iD_\mu 
H)(\overline{E}\gamma^\mu e)$, can be negligible. 
This happens for example when the lightest vector resonance at the scale 
$\Lambda$ is the one associated to $U(1)_X$ in the minimal 
CHM~\cite{Agashe:2004rs} $SO(5)\times U(1)_X/SO(4)\times U(1)_X$, in which case 
it does not interact with the Higgs degrees of freedom before EW symmetry 
breaking~\cite{Panico:2015jxa}. For vector triplets of $SO(4)$, which do 
couple to the Goldstone bosons, the operator $(H^\dagger iD_\mu 
H)(\overline{E}\gamma^\mu e)$ arises with strength $\sim g_*^2 
\rm{s_R}/\Lambda^2$. We have checked that, in this case, the decay via 
effective operators still dominates provided $0.6$ TeV $\lesssim m_E\lesssim 1$ 
TeV.}

Other scenarios that can be captured by the aforementioned EFT include 
$U(1)'$ extensions of the SM gauge group,
in which a new vector boson $V$ gets a mass from a hidden Higgs not charged 
under the SM. It is well known that new VLLs must be generally present to avoid 
gauge anomalies~\cite{Duerr:2013dza,Chao:2015nsm,Chala:2015cev}. The mass of the 
latter is controlled by the Yukawas to the hidden Higgs and so it can be easily 
much smaller than the mass of the $Z'$.

In any case, in this paper we adopt an agnostic EFT approach to the physics 
of $E$.
A thorough inspection of the experimental literature reveals that the search 
of Ref.~\cite{Sirunyan:2020awe}, originally conceived for excited leptons,
might be used to test this scenario. There are however severe limitations to translate the bounds obtained in that paper to our framework. To start
with, only one dimension-six operator is considered in that experimental 
analysis. 
Second, it only considers the decay $E\to \ell q \overline{q}$, with $q$ 
being a light quark, neither a $b$ nor a top. And third, the 
bounds obtained in that search can not be translated to UV models with cut-off 
below $10$ TeV. In order to overcome these weaknesses, 
we recast the experimental analysis in full detail and 
apply it to the entire EFT, for the different decay channels of $E$ 
in a wide range of masses, while keeping strict control of the EFT validity.

The article is organised as follows. In section~\ref{sec:theory} we introduce 
the EFT for the SM extended with $E$ (ESMEFT), 
and discuss its effects on single $E$ production and the subsequent decay. 
In section~\ref{sec:analysis} we recast the most up-to-date search for excited 
leptons and analyse the impact of the different effective operators involving 
$E$  on its production and decay. We derive master formulae that
can be used to automatically predict the number of events expected in any of 
the signal regions of the experimental search for \textit{arbitrary} 
combinations of operators (all of which produce $E$ at very different regions of 
the phase space).
We discuss the validity of the EFT and derive global bounds on the Wilson 
coefficients of the EFT accordingly.
In section~\ref{sec:improvements} we discuss modifications of the current 
analysis that improve the sensitivity to the ESMEFT, at current and future 
luminosities. 
In section~\ref{sec:app} we discuss different UV completions of the ESMEFT, 
particularly all those extending the SM$+E$ renormalizable Lagrangian with just 
one single field, and apply our analyses to constrain their parameter 
spaces. We conclude in section~\ref{sec:conclusions}. We dedicate 
appendix~\ref{app:unitarity} to a discussion of the technical details on the perturbative unitarity limits that we use when studying the validity of the EFT.

\section{Theoretical setup}
\label{sec:theory}
%
We extend the SM with an $SU(2)_L$ singlet VLL $E= E_R+E_L$ with hypercharge 
$Y=-1$. The leading (renormalizable) Lagrangian reads
\begin{equation}
 L = \overline{E}\left(i\slashed{D}-M_E\right)E - \left(y\overline{l_L}H E + \text{h.c.}\right)~.
 \label{eq:yukawa}
 \end{equation}
At dimension six, the following contact interactions contribute to $p p \to E 
\ell$~\footnote{Let us note 
that, within the context of CHMs, the Wilson coefficients $f_{qdl}$, 
$f_{qul}$ and $f_{luq}$ are expected to vanish for $y\to 0$, because either 
the LH or 
the RH SM fermions should be fully elementary in this case. In general, SM four-fermion operators, $(qq)(\ell\ell)$ and $(qq)(qq)$ are also present. Bounds 
on these are comparable~\cite{deBlas:2013qqa,Falkowski:2017pss,Greljo:2017vvb} 
or weaker~\cite{Domenech:2012ai} than those that we obtain 
below for the ESMEFT; but the corresponding operators are suppressed by one more power of the
lepton degree of compositeness.}:
\begin{align}\nonumber
 L  &=  f_{ue} \left(\overline{u_R}\gamma^\mu u_R\right)\left(\overline{e_R} \gamma_\mu E\right) +  
 f_{de} \left(\overline{d_R}\gamma^\mu d_R\right)\left(\overline{e_R}\gamma_\mu E\right)\\\nonumber
 &+ f_{qe} 
 \left(\overline{q_L}\gamma^\mu q_L\right)\left(\overline{e_R}\gamma_\mu  E\right)  
 + f_{qdl}\left(\overline{q_L}d_R\right)\left(\overline{E}l_L\right) \\
 &  + f_{qul} \left(\overline{q_L} u_R\right)\epsilon\left(\overline{l_L}^T E\right) + 
f_{luq}\left(\overline{l_L}u_R\right)\epsilon\left(\overline{q_L}^T E\right) + \text{h.c.}~,
 \label{eq:singleProdOps1}
\end{align}
where $f_i \equiv c_i/\Lambda^2$. As usual, $e_R$ denotes the SM lepton 
singlet; and $u_R$ and $d_R$ and $q_L$ represent the SM singlet quarks and the LH doublet, respectively. We also define 
$\epsilon = i\sigma_2$, with $\sigma_2$ being the second Pauli matrix.  

Remarkably, non four-fermion interactions lead to processes suppressed 
by loop or Yukawa factors or do not grow with energy; and they can therefore be 
neglected. (Evidently, although our research has been triggered by previous 
studies of CHMs, this EFT describes any new physics scenario involving such VLL,
irrespectively of whether any other new physics is much heavier or not; 
it is hence more generic than the usual approach to the phenomenology of VLLs.)

The relations between interaction eigenstates $e$, $E$ 
and the mass eigenstates $e^-$, $E^-$ read
\begin{equation}
 \begin{aligned}
	e_R &= ~~\cos\theta_R e^-_R + \sin\theta_R E^-_R~, \\
	E_R &= -\sin\theta_R e^-_R + \cos\theta_R E^-_R~,
 \end{aligned}
\end{equation}
\begin{equation}
 \begin{aligned}
	e_L &= ~~\cos\theta_L e^-_L + \sin\theta_L E^-_L~, \\
	E_L &= -\sin\theta_L e^-_L + \cos\theta_L E^-_L~,
 \end{aligned}
 \label{eq:rel1}
\end{equation}
for the right and left chiral fields, respectively, where 
\begin{equation}
\sin\theta_L \rightarrow \frac{y v}{\sqrt{2} m_E}~,
\qquad 
\sin\theta_R \rightarrow 0~,
\label{eq:rel2}
\end{equation}
for $y \ll 1$ and in the limit $m_e\rightarrow 0$.
The relation between $M_E$ and the physical mass $m_E$ reads
\begin{equation}
 M_E = \sqrt{m_E^2- \frac{y^2v^2}{2}}~,
 \label{eq:rel3}
\end{equation}
again, in the same limit. 
In what follows we shall denote $\cos\theta_L$ and $\sin\theta_L$ 
by $c_L$ and $s_L$, respectively.

The mixing between the SM charged leptons and $E$ 
modifies the coupling of the $Z$ boson to the left current:
\begin{equation}
 \frac{e}{s_W c_W} g_L \overline{e_L} \gamma^\mu e_L Z_\mu= 
\frac{e}{s_W c_W}\left(g_L^{SM}+\delta g_L\right) \overline{e_L} \gamma^\mu e_L Z_\mu~,
\end{equation}
where $g_L^{SM}$ is the corresponding coupling in the SM, 
$\delta g_L = (y v/\sqrt{2} m_E)^2/2$, 
and $s_W$ and $c_W$ are the sine and 
cosine of the Weinberg angle, respectively.
EWPD provide the following constraint on the mixing 
between the SM fermions and the new heavy VLL 
at the 95\% CL~\cite{deBlas:2013gla}:
\begin{equation}
 |s_L|=\left|\frac{y v}{\sqrt{2}m_E} \right| < 0.021~(0.030)~,
 \label{eq:EPWD}
\end{equation}
for $E$ mixing with electrons (muons). Taking for reference $m_E=0.5$ TeV, 
the bound on $y$ then reads
\begin{equation}
 |y|<0.06~(0.09)~.
 \label{eq:EPWD1}
\end{equation}
The regime $y\ll 1$ is therefore justified. The usual regime in 
which effective 
operators are ignored corresponds to $\Lambda\to\infty$. In both cases, the 
single production cross 
section triggered by $u\overline{u}$, 
to leading order in $\theta_W$ and having neglected $m_Z \ll \sqrt{s}$, reads
\begin{widetext}
\begin{align}
 \frac{d\sigma}{d\theta} = \frac{\sin{\theta}}{32\pi s}\left(1-\frac{m_E^2}{s}\right)\Bigg\lbrace&-\frac{\pi^2\alpha^2}{3s_W^4 s^2}s_L^2 \left(s+t\right) \left(m_E^2-s-t\right) + \frac{1}{3\Lambda^4}\bigg[s\left(s-m_E^2\right)\left(\frac{c_{qul}^2}{4}+c_{ue}^2\right)\nonumber\\
 &+t\left(t-m_E^2\right)\left(\frac{c_{luq}^2}{4}+c_{ue}^2+c_{qe}^2\right) + st \left(2c_{ue}^2-\frac{1}{2}c_{qul}c_{luq}\right)\bigg]\Bigg\rbrace~.
 \label{eq:dXsecSPuu}
\end{align}
\end{widetext}%
Likewise, for the counterpart driven by 
$d\overline{d}$ annihilation, we have the following result for the 
differential cross section:
\begin{widetext}
\begin{align}
 \frac{d\sigma}{d\theta} = \frac{\sin{\theta}}{32\pi s}\left(1-\frac{m_E^2}{s}\right)\Bigg\lbrace&-\frac{\pi^2\alpha^2}{3s_W^4 s^2}s_L^2 \left(s+t\right) \left(m_E^2-s-t\right) + \frac{1}{3\Lambda^4}\bigg[s\left(s-m_E^2\right)\left(\frac{c_{qdl}^2}{4}+c_{de}^2\right)\nonumber\\
 &+t\left(t-m_E^2\right)\left(c_{qe}^2+c_{de}^2\right) + 2st c_{de}^2\bigg]\Bigg\rbrace\,.
 \label{eq:dXsecSPdd}
\end{align}
\end{widetext}%
Integration over $\theta$ can be performed by noticing that 
$t = m_E^2 - 2 p_i \left(\sqrt{m_E^2 + p_f^2} - p_f \cos{\theta}\right)$, 
with $p_i = \sqrt{s}/2$ and $p_f = \left(s - m_E^2\right)/\left(2\sqrt{s}\right)$.

\begin{figure}[h]
 \includegraphics[width=\columnwidth]{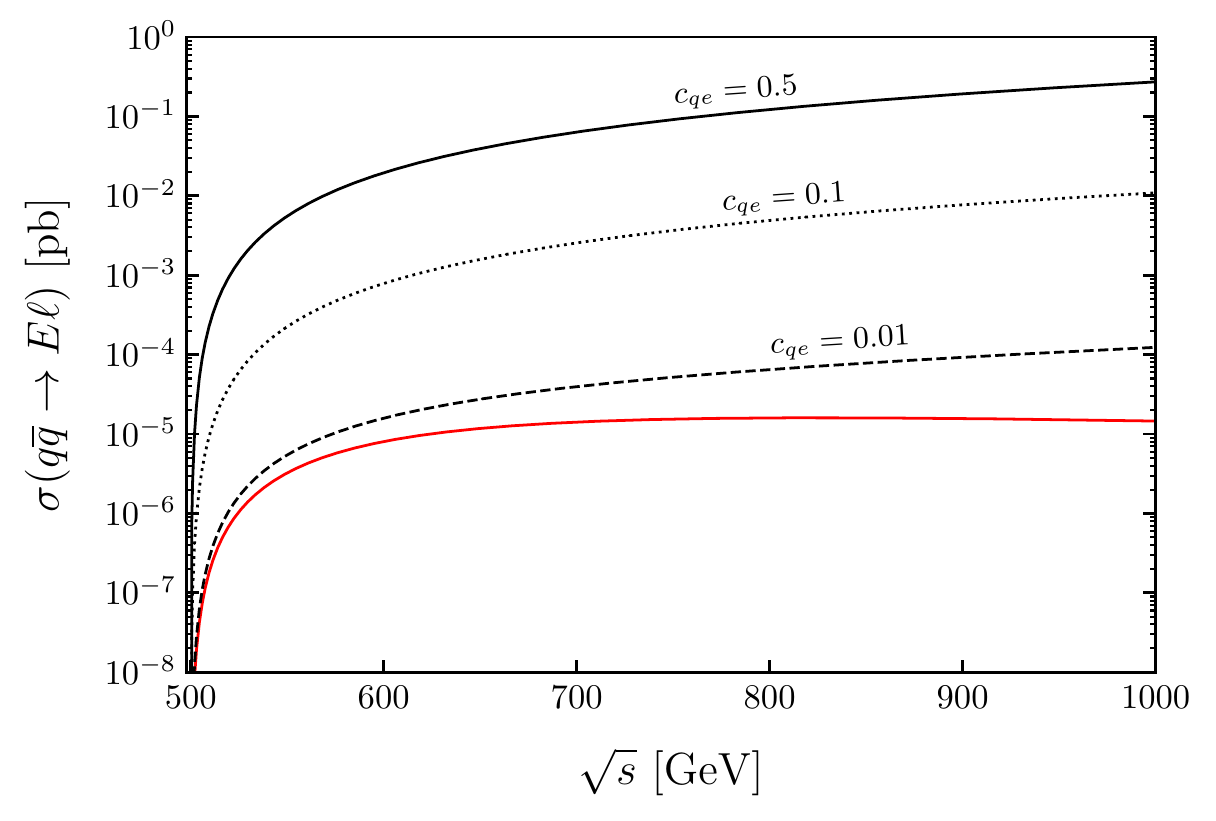} 
 \caption{\it Single production cross section as a function of $\sqrt{s}$, 
 for $m_E = 500$~GeV and $y = 0.1$. 
 The black lines correspond to different values of $c_{qe}$, 
 with $\Lambda = 1$~TeV. All other Wilson coefficients are set to zero.  
 The red line shows the cross section for the SM mediated process.}
 \label{fig:EFTvsSM}
\end{figure}

In Fig.~\ref{fig:EFTvsSM} we present the total single production cross 
section for fixed values of the Wilson coefficient $f_{qe}$ and assuming the 
maximum experimentally allowed value for $y$. 
For comparison, the red line shows  the cross section for $\Lambda\to\infty$, 
in which the only contribution comes from the $Z$ exchange.
This $s$-channel contribution suppressed by the gauge 
boson propagator scales as $\sigma\sim 1/s$. 
On the contrary, in the EFT, 
$\sigma\sim s/\Lambda^4$ and therefore the cross section grows with 
the energy. 
Together with the $y$ suppression, this effect makes the 
effective interactions dominate the cross section in 
the large $\sqrt{s}$ region, even for $f_{qe}$ 
as small as $\sim 0.01$~TeV$^{-2}$. 

The Yukawa coupling $y$ in Eq.~\eqref{eq:yukawa} triggers also
the two-body decay of $E$ into SM gauge bosons, 
$E\to Z/h\ell$ and $E\to W\nu$. 
For $y \ll 1$ and $m_Z \ll m_E$, 
the decay width reads
\begin{equation}
 \Gamma =  y^2 \frac{\alpha}{16 s_W^2 c_W^2} 
 \left(\frac{v}{m_Z}\right)^2 m_E~.
 \label{eq:2body}
\end{equation}

Concerning the three-body decay of $E$, let us first note that, if its interactions are flavour universal, then it couples equally to all
quarks and leptons, and therefore $E$ decays mostly into $\ell q\overline{q}$, 
with $q$ being either a light, a bottom or a top quark; because there are three 
(colour) copies of each quark.
Likewise, if similarly to the Higgs
boson, $E$ couples hierarchically to all fermions according to their masses, 
then its decays to three leptons is again sub-dominant. In light of this
observation, we will neglect the mode $E\to\ell\ell\overline{\ell}$ hereafter. 
This implies that the operators relevant for analysing the decay of $E$ are
also precisely those in Eq.~\eqref{eq:singleProdOps1}.
The differential decay width for $E\to\ell u\overline{u}$ reads 
\begin{widetext}
\begin{align}
 \frac{d\Gamma'}{dE_1 dE_2}= \frac{3}{128\pi^3 m_E}  
 \bigg[&2 E_1 m_E \left(m_E^2 - 2 E_1 m_E\right) 
 \left(f_{luq}^2+4 f_{qe}^2\right) + 2 f_{luq} f_{qul} 
 \left(m_E^2-2 E_1 m_E\right) \left(m_E^2 - 2 E_3 m_E\right) \nonumber \\
 &+2 E_3 f_{qul}^2 m_E \left(m_E^2 - 2 E_3 m_E\right) 
 + 8 E_2 f_{ue}^2 m_E \left(m_E^2 - 2 E_2 m_E\right)\bigg]~.
 \label{eq:3body1}
\end{align}
\end{widetext}%
Analogously, for $E\to\ell d\overline{d}$  we have
\begin{widetext}
\begin{equation} 
 \frac{d\Gamma'}{dE_1 dE_2}= \frac{3}{128\pi^3 m_E}  
 \bigg[8 E_1 f_{qe}^2 m_E \left(m_E^2 -2 E_1 m_E\right) + 
 8 E_2 f_{de}^2 m_E \left(m_E^2 -2 E_2 m_E\right)  + 
 2 E_3 f_{qdl}^2 m_E \left(m_E^2 -2 E_3 m_E\right)\bigg]~,
 \label{eq:3body}
\end{equation}
\end{widetext}%
where $E_1, E_2 $ and $E_3$ are the energies of 
$u(d)$, $\overline{u}(\overline{d})$ and $\ell$, respectively.
Upon integrating over the whole phase space, we arrive at
\begin{align}\nonumber
 \Gamma' = \frac{m_E^5}{2048 \pi^3} \bigg[&f_{luq}^2 +  f_{luq} f_{qul}  + 
f_{qul}^2  + f_{qdl}^2\\
  &+ 4 \left(2f_{qe}^2 + f_{ue}^2+f_{de}^2\right)\bigg]~.
\end{align}

Assuming $\mathcal{O}(1)$ couplings and all quarks, the comparison between 
$\Gamma'$ and $\Gamma$ reveals that the three-body decay dominates for 
$y\lesssim 0.2 \left(m_E/\Lambda\right)^2$. Namely, 
$y\lesssim 0.008~(0.02)$ for $\Lambda/m_E\sim 5\, (3)$. 
This value of $y$ is very close to the EWPD bound, 
it is therefore very likely that $E$ decays predominantly via EFT operators.
Hereafter we study the regime $y\to 0$, and focus only on the case 
$\ell=\mu$. Departures from this assumption are discussed in 
section~\ref{sec:conclusions}.

\section{Collider signatures}
\label{sec:analysis}
\begin{figure}[t]
 \includegraphics[width=\columnwidth]{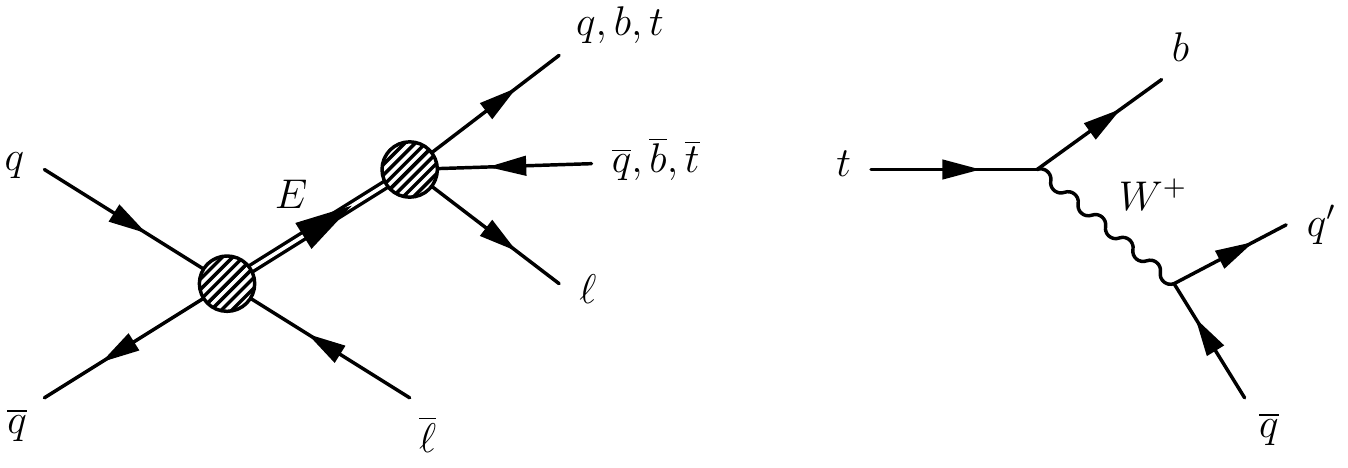} 
 \caption{\it Left: single production of $E$ and its subsequent decay 
 via four-fermion operators. 
 Right: hadronic decay of the top in the SM.}
 \label{fig:diagrams}
\end{figure}

%
In the regime $y\to 0$, the single production of $E$ and its subsequent decays 
proceed as depicted in Fig.~\ref{fig:diagrams}.
The experimental analysis of Ref.~\cite{Sirunyan:2020awe} is optimised for 
the light quark channel, shown in the left panel (the one with $q\overline{q}$).

In general terms, it requires first two isolated leptons
with $p_T > 35$ GeV ($25$ GeV) and $|\eta| <1.44$ or $1.56<|\eta|<2.50$ 
($|\eta|<2.4$) for electrons (muons). (Isolation is defined by the requirement 
that the sum of the $p_T$ of all tracks within $\Delta R =0.3$ of a lepton 
is smaller than 5 GeV.) Likewise, it requires at least two anti-k$_t$ ($R=0.4$) 
jets with $p_T > 50$ GeV. The leading lepton is also required to have $p_T > 
230$ GeV ($53$ GeV) for electrons (muons). Finally, the invariant mass of the 
two leptons must be above $500$ GeV.

The discriminating variable is the invariant mass of the two leptons and the two 
leading jets, $m_{\ell\ell j j}$. It is split into five energy bins: 
$[0.5-1.5]$~TeV, $[1.5-2.5]$~TeV, $[2.5-3.5]$~TeV, $[3.5-4.5]$~TeV, 
$[4.5-10]$~TeV.

In order to determine limits on $f$, we use the energy bin $[1.5-2.5]$~TeV, 
so that our EFT can be used to describe a wide range of UV models. 
(If we use all bins, models with $\Lambda < 10$ TeV 
can not be studied using the EFT approach.)	
Within this energy region, even $f$ of order $\mathcal{O}(1)$~TeV$^{-2}$ 
are allowed by perturbative unitarity constraints; see appendix~\ref{app:unitarity} 
($m_{\ell\ell j j}$ can be used as a proxy for 
the partonic centre-of-mass energy $\sqrt{\hat{s}}$).

Following Eqs.~\eqref{eq:dXsecSPuu} and \eqref{eq:dXsecSPdd}, 
the cross section, and therefore the number
of events in each of these bins, can be written as
\begin{align}\nonumber
 N = \frac{1}{\Lambda^4} 
 \bigg[&\mathcal{I}_1^u \left(\frac{c_{qul}^2}{4}+c_{ue}^2\right)
 + \mathcal{I}_2^u \left(\frac{c_{luq}^2}{4}+c_{ue}^2+c_{qe}^2\right)\\\nonumber
 &+ \mathcal{I}_3^u \left(2c_{ue}^2
 -\frac{1}{2}c_{qul}c_{luq}\right)
 +\mathcal{I}_1^{d}	\left(\frac{c_{qdl}^2}{4}+c_{de}^2\right)\\
 &+\mathcal{I}_2^{d}\left(c_{de}^2+c_{qe}^2\right)+ 2 
 \mathcal{I}_3^{d}c_{de}^2\bigg]~,
 \label{eq:mastereq}
\end{align}
where the coefficients $\mathcal{I}^q_i$, $q = u,d$, $i=1,2,3$, are bin as 
well as mass dependent and must be obtained from simulation. To this aim we 
have generated signal events using \texttt{MadGraph v5}~\cite{Alwall:2014hca} and 
\texttt{Pythia v8}~\cite{Sjostrand:2014zea} for the three cases: 
$E\to\ell q\overline{q}$, $\ell b\overline{b}$, $\ell t\overline{t}$.
To extract $\mathcal{I}^u_1$, $\mathcal{I}^u_2$ and $\mathcal{I}^u_3$, 
we turn on $c_{qul}$, $c_{luq}$ and $c_{ue}$, respectively. 
Furthermore, we set $c_{de} \neq 0$ to realise the decay of $E$ to
the down-type quarks, whereas the semi-leptonic decay of $E$ to 
a pair of tops is triggered by the operator responsible for the production of $E$. 
All other operator coefficients are set to zero.
To obtain $\mathcal{I}^d_1$, $\mathcal{I}^d_2$ and $\mathcal{I}^d_3$, 
we turn on $c_{qdl}$, $c_{qe}$ and $c_{de}$, respectively. 
The same operators trigger the decay of $E$ to the down-type quarks. 
To allow for the decay of $E$ to a pair of tops, we switch on $c_{ue}$, 
except for the second case, 
when $c_{qe} \neq 0$ already ensures such a decay.

The Monte Carlo events are subsequently passed through a recast version of the 
experimental analysis that we have implemented using dedicated routines 
based on \texttt{Fastjet v3}~\cite{Cacciari:2011ma} and 
\texttt{ROOT v6}~\cite{Brun:1997pa,Antcheva:2009zz}. 
We do not include detector simulation. 
We have validated the analysis using the dominant background 
given by Drell-Yan production merged up to two extra matrix element partons, 
finding good agreement with the numbers provided in 
Ref.~\cite{Sirunyan:2020awe} (see Fig.~7 therein).

The coefficients $\mathcal{I}^q_i$ obtained in the way described above 
are shown in Tabs.~\ref{tab:500GeV}, \ref{tab:700GeV} and \ref{tab:900GeV} 
for $m_E = 500$, $700$ and $900$~GeV, respectively. 
We focus on $\ell = \mu$;  the (small) differences for electrons due to the 
different detector response are succinctly discussed in section~\ref{sec:conclusions}. 
\begin{table*}[t]
 \renewcommand{\arraystretch}{1.2}
 \centering
 \begin{tabular}{|c|c|ccccc|}
  \hline
  \multicolumn{2}{|c|}{} & \multicolumn{5}{|c|}{Bins in $2\ell2j$ mass [TeV]} \\
  \multicolumn{2}{|c|}{} & $0.5-1.5$ & $1.5-2.5$ & $2.5-3.5$ & $3.5-4.5$ & $4.5-10$\\
  \hline
  \hline
  \multirow{3}{*}{\vtext{$E \to \mu d \overline{d}$}} & $\mathcal{I}^q_1/10^{2}$ 
  & 390 (150) & 530 (240) & 220 (96) & 62 (19) & 22 (5.5)  \\
  & $\mathcal{I}^q_2/10^{2}$ & 140 (88) & 180 (100) & 74 (35) & 19 (9.2) & 8.4 (2.9)   \\
  & $\mathcal{I}^q_3/10^{2}$ & $-190~(-73)$ & $-260~(-120)$ & $-110~(-49)$ & $-29~(-9.1)$ & $-11~(-2.8)$ \\ 
  \hline
  \hline
  \multirow{3}{*}{\vtext{$E \to \mu b \overline{b}$}} & $\mathcal{I}^q_1/10^{2}$ 
& 380 (150) & 480 (210) & 210 (88) & 55 (22) & 18 (3.5)  \\
  & $\mathcal{I}^q_2/10^{2}$ & 140 (85) & 170 (97) & 68 (33) & 22 (7.4) & 5.5 (2.1)   \\
  & $\mathcal{I}^q_3/10^{2}$ & $-190~(-79)$ & $-240~(-110)$ & $-100~(-44)$ & $-29~(-11)$ & $-9.2~(-1.7)$ \\ 
  \hline
  \hline
  \multirow{3}{*}{\vtext{$E \to \mu t \overline{t}$}} & $\mathcal{I}^q_1/10^{2}$ 
& 170 (160) & 200 (120) & 73 (27) & 20 (5.7) & 4.3 (1.4)  \\
  & $\mathcal{I}^q_2/10^{2}$ & 93 (57) & 85 (48) & 24 (9.6) & 5.3 (1.4) & 1.0 (0.36)   \\
  & $\mathcal{I}^q_3/10^{2}$ & $-82~(-79)$ & $-110~(-64)$ & $-37~(-13)$ & $-11~(-2.5)$ & $-2.1~(-0.53)$ \\ 
  \hline
  \hline
  \multicolumn{2}{|c|}{SM} & $949\pm 115$ & $161\pm25$ & $13.7\pm 3.7$  & $1.2\pm 0.6$ & $0.48\pm 0.32$ \\
  \multicolumn{2}{|c|}{Data} & $949$ & $151$ & $11$ & $0$ & $1$ \\
  \multicolumn{2}{|c|}{$s_\mathrm{max}$} & $291$ & $60$ & $14$ & $4$ & $5$ \\
  \hline
 \end{tabular}
 \caption{\it Coefficients $\mathcal{I}^q_i$, $q=u~(d)$, in TeV$^4$ 
 and rounded to two significant figures
 for $pp\rightarrow \mu^+ \mu^{-} jj$
 obtained upon recasting the experimental analysis of 
 Ref.~\cite{Sirunyan:2020awe} for $\sqrt{s} = 13$~TeV 
 and total integrated luminosity $\mathcal{L} = 77.4$~fb$^{-1}$. 
 We have assumed $m_E = 500$~GeV and 
 $\mathcal{B}(E \to \mu d \overline{d}) = 1$ (top), 
 $\mathcal{B}(E \to \mu b \overline{b}) = 1$ (middle), and 
 $\mathcal{B}(E \to \mu t \overline{t}) = 1$ (bottom). 
 We also display the SM prediction, the data 
 and the maximal allowed signal $s_\mathrm{max}$ in each bin 
 (for muons in the final state). 
 This latter number is computed using the CL$_s$ method, 
 taking into account the uncertainty on the background 
 displayed in the table as well as 15\% uncertainty on the signal; 
 see the text for details.}
 \label{tab:500GeV}
\end{table*}
\begin{table*}[!h]
 \renewcommand{\arraystretch}{1.2}
 \centering
 \begin{tabular}{|c|c|ccccc|}
  \hline
  \multicolumn{2}{|c|}{} & \multicolumn{5}{|c|}{Bins in $2\ell2j$ mass [TeV]} \\
  \multicolumn{2}{|c|}{} & $0.5-1.5$ & $1.5-2.5$ & $2.5-3.5$ & $3.5-4.5$ & $4.5-10$\\
  \hline
  \hline
  \multirow{3}{*}{\vtext{$E \to \mu d \overline{d}$}} & $\mathcal{I}^q_1/10^{2}$ 
& 230 (79) & 480 (210) & 230 (100) & 73 (30) & 26 (7.5)  \\
  & $\mathcal{I}^q_2/10^{2}$ & 89 (57) & 170 (96) & 78 (39) & 24 (10) & 7.9 (2.0)   \\
  & $\mathcal{I}^q_3/10^{2}$ & $-110~(-40)$ & $-240~(-110)$ & $-110~(-51)$ & $-35~(-15)$ & $-12~(-3.4)$ \\ 
  \hline
  \hline
  \multirow{3}{*}{\vtext{$E \to \mu b \overline{b}$}} & $\mathcal{I}^q_1/10^{2}$ 
& 260 (95) & 460 (210) & 200 (86) & 74 (26) & 21 (4.9)  \\
  & $\mathcal{I}^q_2/10^{2}$ & 94 (58) & 160 (86) & 68 (36) & 20 (9.1) & 6.5 (2.0)   \\
  & $\mathcal{I}^q_3/10^{2}$ & $-130~(-47)$ & $-230~(-100)$ & $-99~(-44)$ & $-36~(-13)$ & $-11~(-2.1)$ \\ 
  \hline  
  \hline
  \multirow{3}{*}{\vtext{$E \to \mu t \overline{t}$}} & $\mathcal{I}^q_1/10^{2}$ 
& 180 (150) & 270 (160) & 100 (50) & 30 (11) & 10 (2.6)  \\
  & $\mathcal{I}^q_2/10^{2}$ & 95 (60) & 110 (57) & 37 (17) & 11 (4.2) & 2.5 (0.53)   \\
  & $\mathcal{I}^q_3/10^{2}$ & $-85~(-74)$ & $-140~(-82)$ & $-52~(-25)$ & $-16~(-5.6)$ & $-5.2~(-1.2)$ \\ 
  \hline
 \end{tabular}
 \caption{\it Coefficients $\mathcal{I}^q_i$, $q=u~(d)$, in TeV$^4$ 
 and rounded to two significant figures
 for  $pp\rightarrow \mu^+ \mu^{-} jj$
 obtained upon recasting the experimental analysis of 
 Ref.~\cite{Sirunyan:2020awe} for $\sqrt{s} = 13$~TeV 
 and total integrated luminosity $\mathcal{L} = 77.4$~fb$^{-1}$. 
 We have assumed $m_E = 700$~GeV and 
 $\mathcal{B}(E \to \mu d \overline{d}) = 1$ (top), 
 $\mathcal{B}(E \to \mu b \overline{b}) = 1$ (middle), and 
 $\mathcal{B}(E \to \mu t \overline{t}) = 1$ (bottom).}
 \label{tab:700GeV}
\end{table*}
\begin{table*}[t]
 \renewcommand{\arraystretch}{1.2}
 \centering
 \begin{tabular}{|c|c|ccccc|}
  \hline
  \multicolumn{2}{|c|}{} & \multicolumn{5}{|c|}{Bins in $2\ell2j$ mass [TeV]} \\
  \multicolumn{2}{|c|}{} & $0.5-1.5$ & $1.5-2.5$ & $2.5-3.5$ & $3.5-4.5$ & $4.5-10$\\
  \hline
  \hline
  \multirow{3}{*}{\vtext{$E \to \mu d \overline{d}$}} & $\mathcal{I}^q_1/10^2$ & 
120 (35) & 400 (170) & 210 (87) & 72 (25) & 29 (8.1) \\
  & $\mathcal{I}^q_2/10^2$ & 43 (27) & 150 (84) & 68 (36) & 25 (11) & 9.7 (3.5) \\
  & $\mathcal{I}^q_3/10^2$ & $-60~(-17)$ & $-200~(-86)$ & $-100~(-43)$ & $-37~(-13)$ & $-15~(-4.0)$ \\ 
  \hline  
  \hline
  \multirow{3}{*}{\vtext{$E \to \mu b \overline{b}$}} & $\mathcal{I}^q_1/10^2$ & 
140 (51) & 380 (170) & 190 (81) & 66 (23) & 23 (5.7) \\
  & $\mathcal{I}^q_2/10^2$ & 52 (33) & 140 (78) & 65 (33) & 22 (8.7) & 7.2 (2.3) \\
  & $\mathcal{I}^q_3/10^2$ & $-68~(-25)$ & $-190~(-86)$ & $-94~(-41)$ & $-33~(-11)$ & $-11~(-2.6)$ \\ 
  \hline  
  \hline
  \multirow{3}{*}{\vtext{$E \to \mu t \overline{t}$}} & $\mathcal{I}^q_1/10^2$ & 
110 (100) & 230 (140) & 100 (50) & 34 (14) & 12 (2.8) \\
  & $\mathcal{I}^q_2/10^2$ & 64 (39) & 100 (53) & 38 (19) & 12 (4.3) & 2.7 (1.1) \\
  & $\mathcal{I}^q_3/10^2$ & $-56~(-52)$ & $-120~(-69)$ & $-50~(-24)$ & $-18~(-6.8)$ & $-5.2~(-1.5)$ \\ 
  \hline
 \end{tabular}
 \caption{\it Coefficients $\mathcal{I}^q_i$, $q = u~(d)$, in TeV$^4$ 
 and rounded to two significant figures
 for  $pp\rightarrow \mu^+ \mu^{-} jj$
 obtained upon recasting the experimental analysis of 
 Ref.~\cite{Sirunyan:2020awe} for $\sqrt{s} = 13$~TeV 
 and total integrated luminosity $\mathcal{L} = 77.4$~fb$^{-1}$. 
 We have assumed $m_E = 900$~GeV and 
 $\mathcal{B}(E \to \mu d \overline{d}) = 1$ (top), 
 $\mathcal{B}(E \to \mu b \overline{b}) = 1$ (middle), and 
 $\mathcal{B}(E \to \mu t \overline{t}) = 1$ (bottom).}
 \label{tab:900GeV}
\end{table*}

Using these tables, we have compared the predicted number of signal events 
in each bin as derived from Eq.~\eqref{eq:mastereq} 
to that obtained directly from simulation for $\mathcal{O}(100)$ different 
combinations of Wilson coefficients. The latter is always contained in a band of 
$\pm 15\%$ around the former~\footnote{Note that in deriving 
Eq.~\eqref{eq:mastereq}, we have neglected the impact of the different effective 
operators triggering the decay of $E$ on the efficiency of the analysis. The 
difference in the efficiencies for selecting single produced events in two 
samples that differ only 
by the operator driving the decay of $E$ is small.  
Moreover, this difference tends to vanish if the two operators are linear 
combinations of $\lbrace\mathcal{O}_{luq},\mathcal{O}_{qe}, 
\mathcal{O}_{ue}\rbrace$ or $\lbrace\mathcal{O}_{de},\mathcal{O}_{qe}\rbrace$. 
The reason is that the differential $E$ decay widths (see 
Eqs.~\eqref{eq:3body1} and \eqref{eq:3body}) driven by two operators within 
the same set differ only by
 $E_1\leftrightarrow E_2$, 
while the cuts are the same for all jets. 
We have checked that this fact 
reflects well on the simulation.}. We therefore take $15\%$ as the systematic 
error in our prediction for the number of signal events. 

We also report in Tab.~\ref{tab:500GeV} the number of observed events as well as the number of expected SM events as given in Ref.~\cite{Sirunyan:2020awe}.  
Using the CL$_s$ method~\cite{Read:2002hq}, 
including the aforementioned $15\%$ uncertainty 
on the signal as well as the uncertainties on the background, 
we derive the maximum number of allowed signal events in each bin. 
These numbers are also shown in the table.

We note that the coefficients $\mathcal{I}$ for light quarks and bottoms are 
roughly equivalent. (We will make use of this observation in what follows 
to derive bounds on the Wilson coefficients as a function of only 
$\mathcal{B}(E\to\ell j\overline{j})\equiv \mathcal{B}(E\to\ell 
q\overline{q})+\mathcal{B}(E\to\ell b\overline{b})= 1-\mathcal{B}(E\to\ell 
t\overline{t})$.) The main reason is that the final states in both cases are 
very similar and no $b$-tagging is at play.

The situation is very different for the top channel. 
The larger number of jets in the final state, 
together with the relatively small top quark leptonic branching ratio, 
makes the corresponding $\mathcal{I}$s even more than a factor of two smaller.

\begin{figure*}[t] 
 \includegraphics[width=\columnwidth]{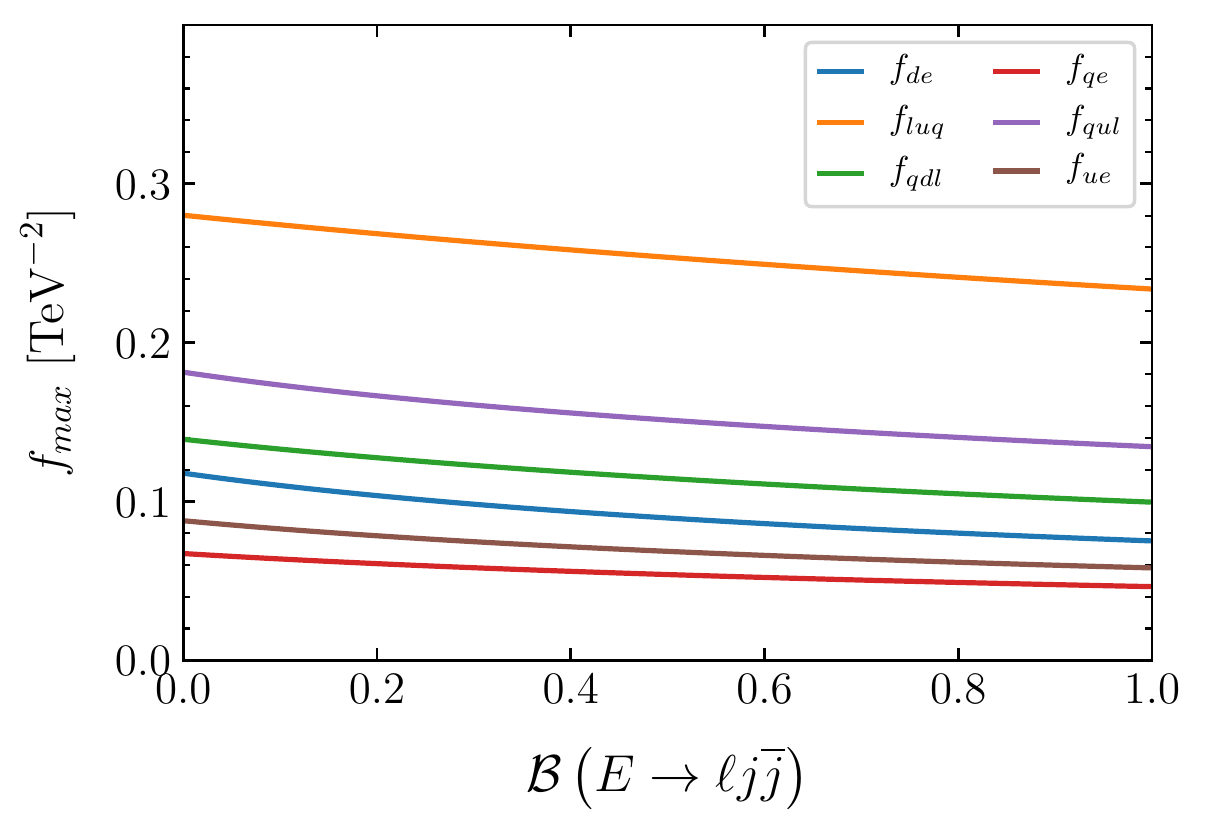} 
 \hspace{0.2cm}
 \includegraphics[width=\columnwidth]{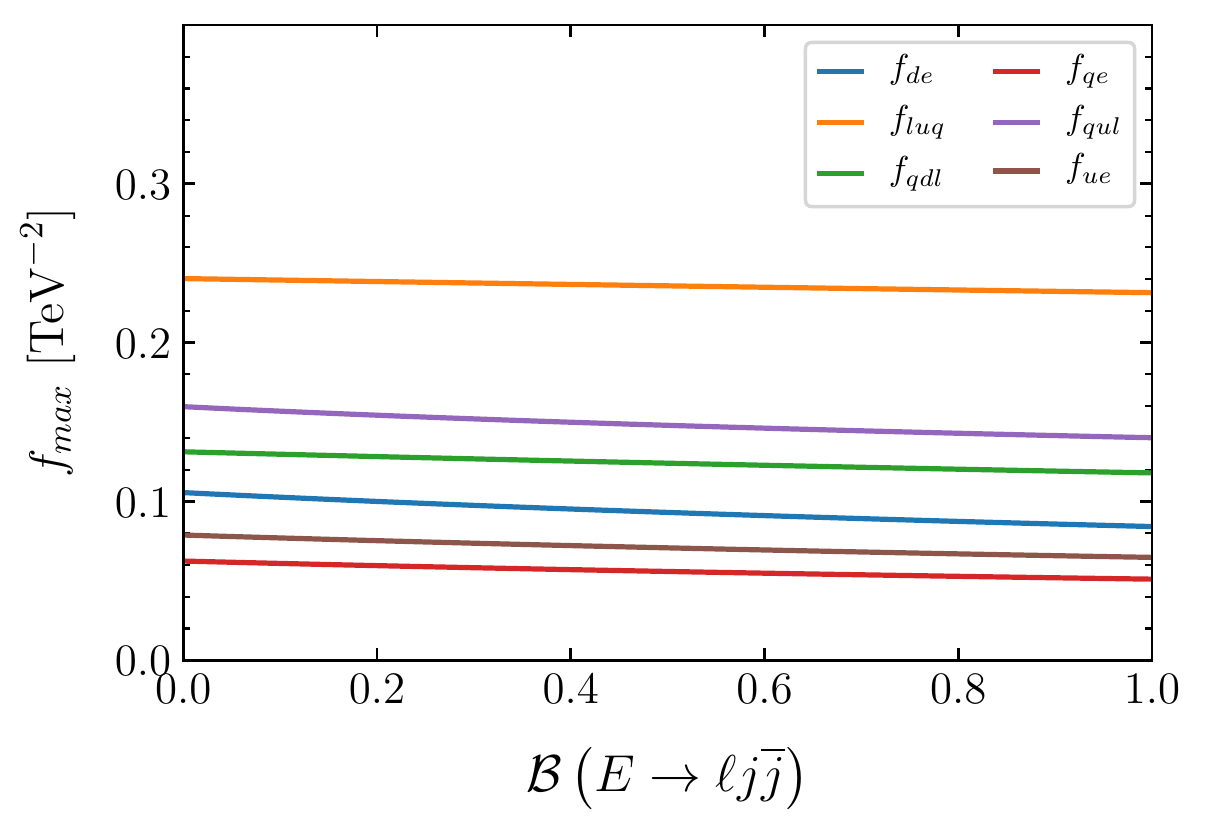} 
 \caption{\it The global limits on the EFT coefficients $f$
 for $m_E=0.5$~TeV (left) and $m_E = 0.9$~TeV (right), 
 using the second bin defined in Tabs.~\ref{tab:500GeV} and \ref{tab:900GeV}.}
 \label{fig:globLim}
\end{figure*}

In Fig.~\ref{fig:globLim} we show the limits on each of the operators of the 
EFT for $m_E=500$ GeV and for $m_E=900$ GeV. 
For setting bounds on $f_{luq}$ we marginalise over $f_{qul}$ 
(as they interfere among themselves); and \textit{vice versa}. 
Note also that for these maximum values of $f$, the energy bin used in the 
analysis, $[1.5-2.5]$ TeV, is well within the energy regime of validity of the 
EFT in light of perturbative unitarity constraints; see 
appendix~\ref{app:unitarity}.

\section{Improvements and prospects}
\label{sec:improvements}
%
Extending the aforementioned experimental analysis with cuts 
on appropriate new observables can make it more sensitive to the ESMEFT. 
One such observable is the invariant mass of the reconstructed $E$.
Note that, because $E$ is heavy, it carries less momentum than the lepton 
in 
$pp\to E \ell$.
Therefore, this lepton is typically the hardest one. 
This effect is strengthened by the fact that 
when $E$ decays it releases energy to several particles.

\begin{figure}[t]
 \includegraphics[width=\columnwidth]{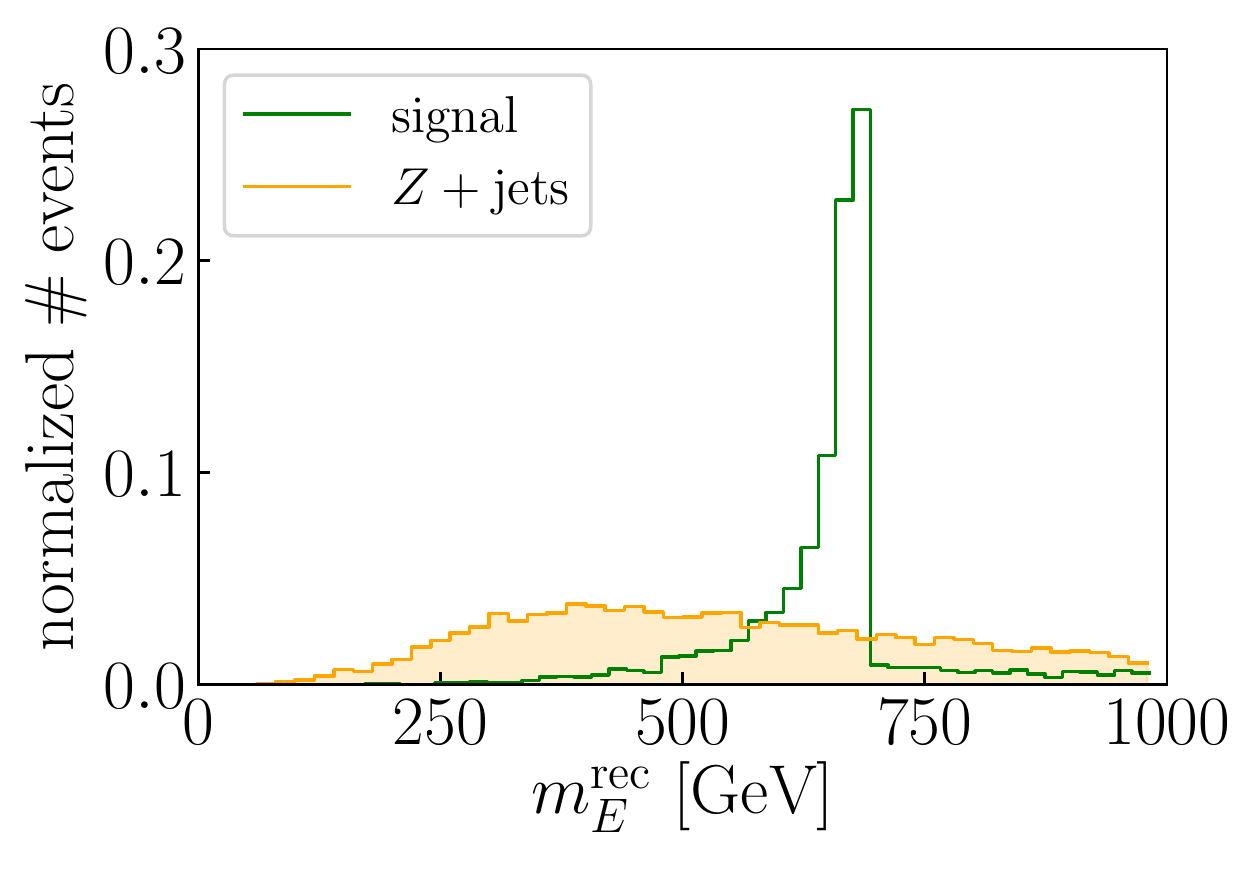}
 \caption{\it Normalized distribution of $m_E^\text{rec}$ right after the cut on 
 $m_{\ell^+\ell^-}>500$ GeV, in the signal for $\mathcal{B}(E\to\mu 
 q\overline{q})=1$ and in the main background.}
 \label{fig:mrecq}
\end{figure}

Thus, one can reconstruct the four-momentum of $E$ as the sum of the 
four-momenta of the softest lepton and the two hardest jets. 
The invariant mass of this object, $m^\text{rec}_E$, peaks well 
around the actual $m_E$ when $\mathcal{B}(E\to \ell q\overline{q}) \sim 1$;
see Fig.~\ref{fig:mrecq}. 
(For this figure we assume $m_E = 700$~GeV. Given the
low sensitivity of our previous results to $m_E$, and because it is in 
between the two extreme cases, $m_E=500$~GeV and $m_E=900$~GeV 
considered before, we restrict to this value hereafter.) 
The main background, ensuing from $Z+\text{jets}$ 
is also shown for comparison.

We extend the current analysis with the extra cut 
$650\,\text{GeV}<m^\text{rec}_E<750\,\text{GeV}$.
In good approximation, the fraction of signal events 
that do not only pass all previous analysis cuts but also this extra one 
is bin and operator independent and of about $0.6$. 
In the background, however, this number goes down to $\sim 0.1$.  

\begin{figure}[t]
 \includegraphics[width=\columnwidth]{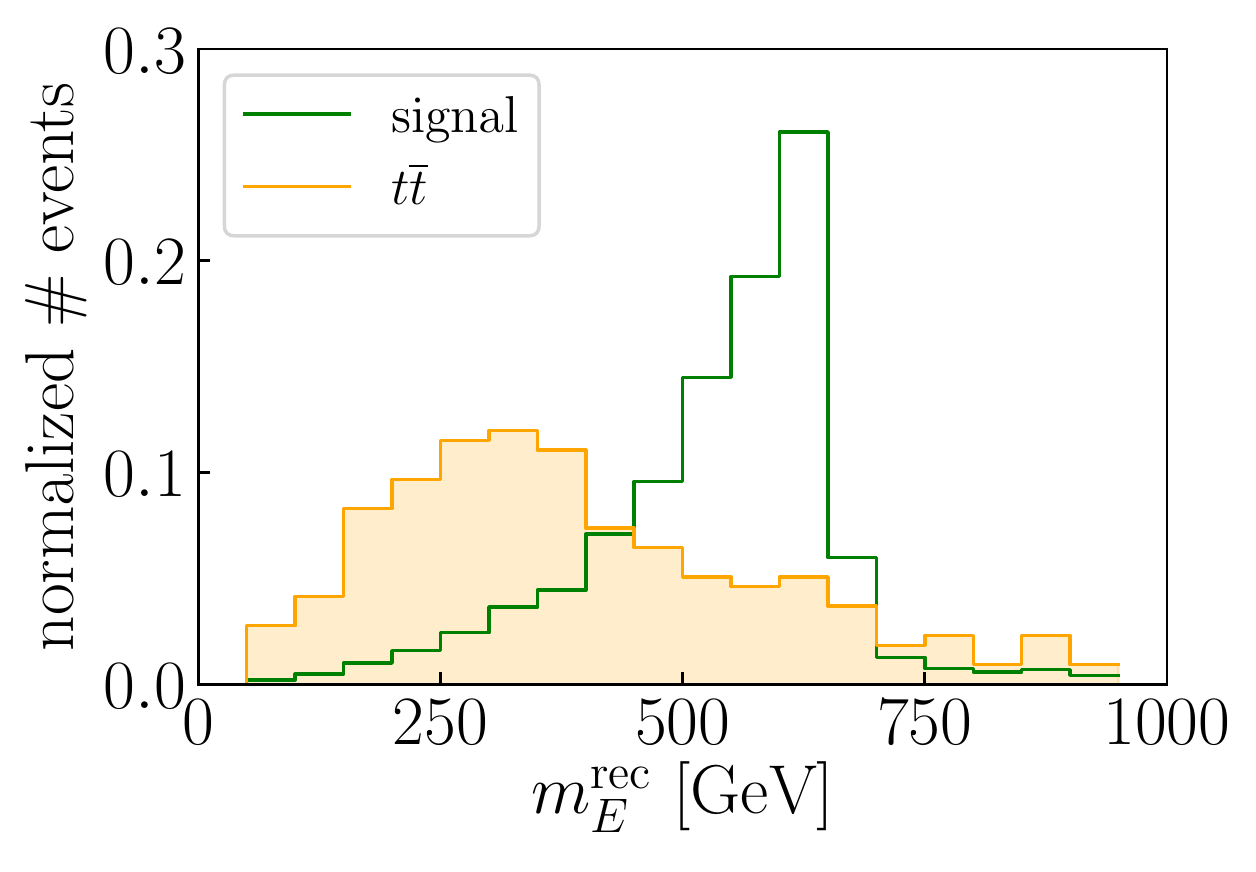}
 \caption{\it Normalized distribution of $m_E^\text{rec}$ right after the cut on 
 $m_{\ell^+\ell^-}>500$ GeV and after requiring exactly two $b$-jets, in the 
 signal for $\mathcal{B}(E\to\mu b\overline{b})=1$ and in the main background.}
 \label{fig:mrecb}
\end{figure}

The search has to be modified in a different way 
if one aims to be more sensitive to the case 
$\mathcal{B}(E\to \ell b\overline{b})\sim 1$ or to  
$\mathcal{B}(E\to \ell t\overline{t})\sim 1$.
In the bottom channel, we require the presence of exactly two $b$-tagged jets. 
(In our simulation, $b$-jet candidates are selected among those jets 
with a $B$-meson within a cone of radius $\Delta R=0.5$; 
the $b$-tagging efficiency is subsequently set to $0.7$.)
We then reconstruct $E$ as the sum of the two leading $b$-jets and the softest
lepton. The invariant mass of the reconstructed $E$ is shown in Fig.~\ref{fig:mrecb} for both the signal and the main background, 
which in this case is $t\overline{t}$
(because the $b$-tagging requirement reduces $Z + \text{jets}$ to 
negligible levels). In this case we require 
$550~\text{GeV} < m_{E}^\text{rec} < 700~\text{GeV}$.
The fraction of signal events surviving the new cuts is $\sim 0.25$, 
while for the background we get $\sim 0.05$.

\begin{figure}[t]
 \includegraphics[width=\columnwidth]{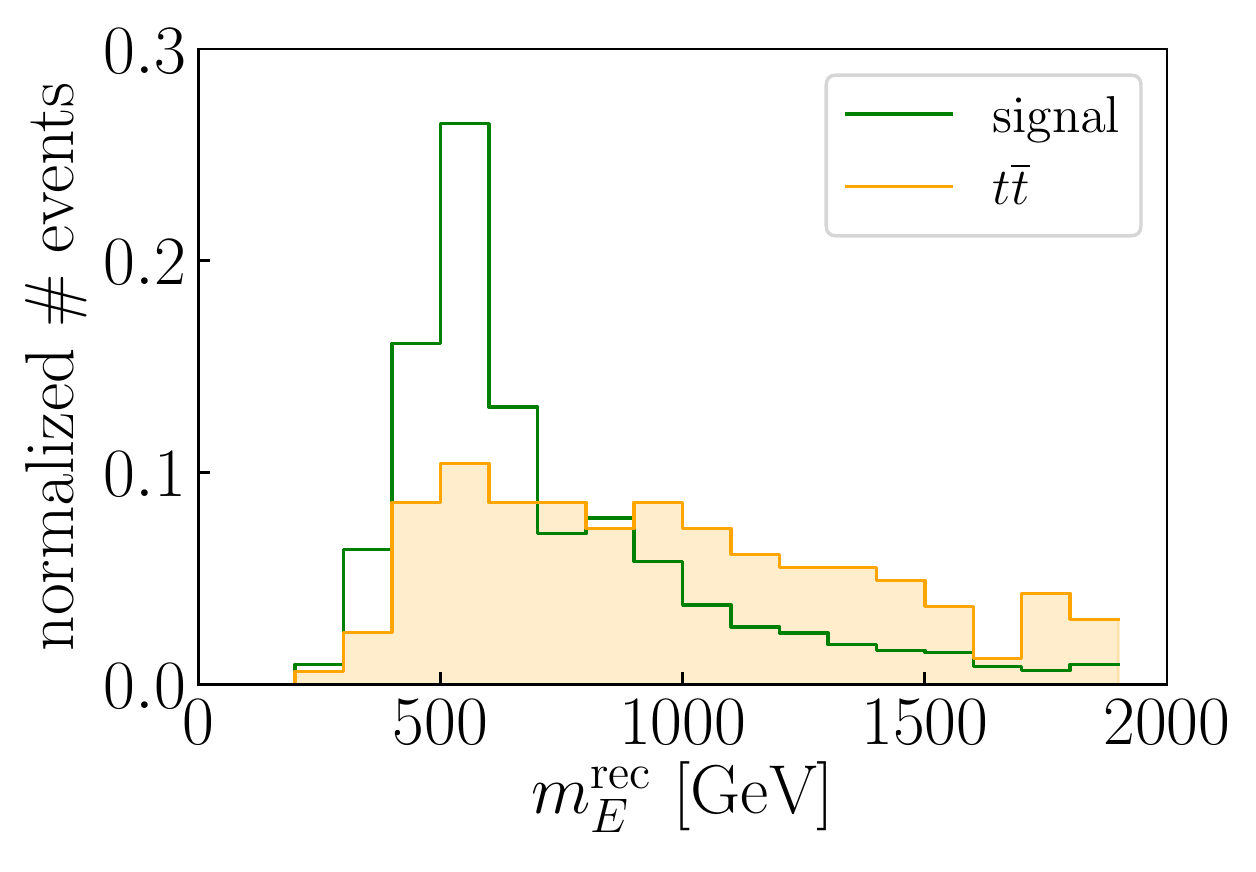}
 \caption{\it Normalized distribution of $m_E^\text{rec}$ right after the cut on 
 $m_{\ell^+\ell^-}>500$ GeV and after requiring exactly two $b$-jets and at 
 least three light jets, in the signal for $\mathcal{B}(E\to\mu t\overline{t})=1$ 
 and in the main background.}
 \label{fig:mrect}
\end{figure}

Finally, in the top channel, in addition to requiring exactly two $b$-jets, 
we demand the presence of at least three light jets. 
We subsequently reconstruct $E$ as the sum of the softest lepton, 
the two $b$-jets and the main three light jets. The corresponding
$m_E^\text{rec}$ is depicted in Fig.~\ref{fig:mrect} in the signal and in 
$t\overline{t}$. 
We require in this case $500~\text{GeV}<m_{E}^\text{rec}<800~\text{GeV}$.
The fraction of signal (background) events surviving the new extra cuts is $\sim 
0.2~(0.05)$.
These numbers reflect the smaller difference between signal and background in 
this case.

Using the CL$_s$ method, assuming again a 15\% uncertainty on the signal 
and the same uncertainties as before for the background, 
and assuming the data to be well described by the SM, 
we obtain the values of $s_\mathrm{max}$ shown in Tab.~\ref{tab:smax}. 
They also include the numbers for the high-luminosity phase of the LHC 
(HL-LHC), 
in which the collected luminosity will reach $\mathcal{L}=3$~ab$^{-1}$.
\begin{table}[t]
 \vspace{0.5cm}
 \centering
 \begin{tabular}{|l|cccc|}
  \hline
  & \multicolumn{4}{|c|}{Bins in $2\ell2j$ mass [TeV]} \\
  & $0.5-1.5$ & 
$1.5-2.5$ & $2.5-3.5$ & $3.5-4.5$ \\
  \hline
  \hline
  $E\to \ell q\overline{q}$ & 46 (549) & 14 (210) & 5 (84) &  4 (68)\\
  $E\to \ell b\overline{b}$ & 14 (210) & 6 (101)  & 4 (68) &  4 (68)\\
  $E\to \ell t\overline{t}$ & 14 (210) & 6 (101)  & 4 (68) &  4 (68)\\
  \hline
 \end{tabular}
 \caption{\it Values of $s_\mathrm{max}$ in four signal regions 
 of the improved analyses with collected luminosity of 
 $\mathcal{L}  =77.4$~fb$^{-1}$ (HL-LHC with $\mathcal{L}=3$~ab$^{-1}$).}
 \label{tab:smax}
\end{table}
Using these numbers, we demonstrate that 
the improved analyses can strengthen the sensitivity on $f$ by more than 
50\%; 
see Tab.~\ref{tab:improvedbounds}. 
\begin{table*}[t]
 \centering
 \begin{tabular}{|l|ccc|}
  \hline
  & $\mathcal{B}(E\to\mu q\overline{q}) =1$ & $\mathcal{B}(E\to\mu 
  b\overline{b})=1$ & $\mathcal{B}(E\to\mu t\overline{t})=1$\\
  \hline
  \hline
  $f_{ue}$ & 0.060, 0.037 (0.023) 		& 0.060, 0.038	(0.025)		 
  & 0.076, 0.050 (0.035)		\\
  $f_{de}$ & 0.079, 0.049 (0.031) 		& 0.081, 0.051 (0.034)		
  & 0.100, 0.072 (0.047)		\\
  $f_{qe}$ & 0.048, 0.030 (0.019)		& 0.049, 0.031 (0.021)		
  & 0.060, 0.042 (0.028)		\\
  $f_{qdl}$ & 0.110, 0.066 (0.041)		& 0.110, 0.067 (0.044)		
  & 0.120, 0.085	(0.056)	\\
  $f_{qul}$ & 0.130, 0.082 (0.051)		& 0.130, 0.083 (0.055)		
  & 0.160, 0.110 (0.072)		\\
  $f_{luq}$ & 0.220, 0.140 (0.086)		& 0.220, 0.140 (0.093)		
  & 0.240, 0.170 (0.110)		\\
  \hline
 \end{tabular}
 \caption{\it Bounds on the Wilson coefficients rounded to two significant 
 figures, in TeV$^{-2}$, in the current and improved (future) analyses. 
 We have assumed $m_E = 700$ GeV and used the energy bin $[1.5-2.5]$~TeV.} 
 \label{tab:improvedbounds}
\end{table*}
%

\section{Applications}
\label{sec:app}
%
The single-field extensions of the SM$+E$ that contribute to the EFT 
at tree level are summarised in Tab.~\ref{tab:extensions}. 
The names of the new scalars follow Ref.~\cite{deBlas:2014mba},
and those of the new vectors Ref.~\cite{delAguila:2010mx}.
\begin{table*}[!t]
 \renewcommand{\arraystretch}{1.2}
 \centering
 \begin{tabular}{|c|c|c|}
  \hline
  Field & Relevant fermionic current & Wilson coefficients\\
  \hline
  \hline
  $\varphi\sim (1,2)_\frac{1}{2}$ & $J= y^E 
  \overline{E} l_L + y^d \overline{d_R} q_L + y^u i\sigma_2 
  \overline{q_L}^T u_R$ & 
  $f_{qdl}=\dfrac{y^d y^E}{m^2}~, ~~
  f_{qul}=-\dfrac{y^u y^E}{m^2} $ \rule[-2ex]{0pt}{6ex} \\[0.2cm]
  $\omega_1\sim (3,1)_{-\frac{1}{3}}$ & $J = y^{Eu} 
  \overline{E^c} u_R + y^{ql} \overline{q_L^c}i\sigma_2 
  l_L+y^{eu} 
  \overline{e_R^c} u_R$ & 
  $f_{ue} = \dfrac{y^{Eu}y^{eu}}{2m^2}~, ~~
  f_{qul}=-\dfrac{y^{Eu}y^{ql}}{m^2}~, ~~ 
  f_{luq}=\dfrac{y^{Eu}y^{ql}}{m^2}$ \\[0.2cm]
  $\omega_4\sim (3,1)_{-\frac{4}{3}}$ & $J = 
  y^{Ed}\overline{E^c}d_R + y^{ed}\overline{e^c_R}d_R$ & 
  $f_{de} = \dfrac{y^{Ed} y^{ed}}{2m^2}$ \\[0.2cm]
  $\Pi_7\sim (3,2)_{\frac{7}{6}}$ & $J = 
  y^{Eq}\overline{E} q_L + y^{lu} i\sigma_2\overline{l_L}^T u_R 
  + y^{eq}\overline{e_R} q_L $ & 
  $f_{qe}=-\dfrac{y^{Eq} y^{eq}}{2m^2}~, ~~ 
  f_{luq}=-\dfrac{y^{Eq}y^{lu}}{m^2}$ \\[0.2cm]
  \hline
  \hline
  $\mathcal{B}_\mu\sim (1,1)_0$ & $J_\mu = 
  g^{E}\overline{e_R}\gamma_\mu E + g^{u} 
  \overline{u_R}\gamma^\mu u_R + g^d\overline{d_R}\gamma^\mu d_R     $& 
  $f_{de} = -\dfrac{g^E g^d}{m^2}~, ~~
  f_{ue}=-\dfrac{g^E g^u}{m^2}~,$ \rule[-2ex]{0pt}{6ex} \\[0.2cm]
  & $+ g^{q} 
  \overline{q_L}\gamma^\mu q_L$  & $f_{qe}=-\dfrac{g^E 
  g^q}{m^2}$ \\[0.2cm]
  $\mathcal{U}_\mu^2\sim (3,1)_{\frac{2}{3}}$ & $J_\mu = 
  g^{Ed}\overline{E}\gamma_\mu d_R + 
  g^{lq}\overline{l_L}\gamma_\mu q_L + 
  g^{ed}\overline{e_R}\gamma_\mu d_R$ & 
  $f_{de} = -\dfrac{g^{Ed} g^{ed}}{m^2}~, ~~
  f_{qdl}=\dfrac{2g^{lq}g^{Ed}}{m^2}$ \\[0.2cm]
  $\mathcal{U}_\mu^5\sim (3,1){\frac{5}{3}}$ & $J_\mu = 
  g^{Eu} \overline{E}\gamma_\mu u_R + 
  g^{eu}\overline{e_R}\gamma_\mu u_R$ & 
  $f_{ue}=-\dfrac{g^{Eu} g^{eu}}{m^2}$ \\[0.2cm]
  $\mathcal{Q}_\mu^5\sim (3,2)_{-\frac{5}{6}}$ & $J_\mu = 
  g^{Eq}\overline{E^c}\gamma_\mu q_L + g^{dl} 
  \overline{d_R^c}\gamma_\mu l_L + 
  g^{eq}\overline{e_R^c}\gamma_\mu q_L$ & 
  $f_{qe}=\dfrac{g^{Eq} g^{eq}}{m^2}~, ~~
  f_{qdl}= -\dfrac{2g^{Eq}g^{dl}}{m^2}$ \\[0.2cm]
  \hline
 \end{tabular}
 \caption{\it The relevant Lagrangian for a scalar $\sigma$ is 
 $L = \partial_\mu\sigma^\dagger\partial^\mu\sigma - 
 m_\sigma^2\sigma^\dagger\sigma - (\sigma^\dagger J_\sigma + \text{h.c.})$.
 For a vector $V$ we have instead 
 $L= -\partial_\mu V_\nu^\dagger\partial^{[\mu} V^{\nu]} + 
 m_V^2 V_\mu^\dagger V^\mu -(V^{\mu\dagger} J_\mu^V+\text{h.c.})$. 
 For each row in the top (bottom) part of the table, $m = m_\sigma~(m_V)$.}
 \label{tab:extensions} 
\end{table*}
In general, more than one EFT operator is generated. 
Assuming $m_E=700$ GeV, we can use Eq.~\eqref{eq:mastereq} 
together with Tab.~\ref{tab:700GeV} and the values of $s_\mathrm{max}$ 
reported in Tabs.~\ref{tab:500GeV} and~\ref{tab:smax}
to derive bounds on the space of couplings for a fixed mass of 
the heavy mediator (set to $5$ TeV), taking \textit{all} operators into account.

\begin{figure*}[t]
 \includegraphics[width=0.5\columnwidth]{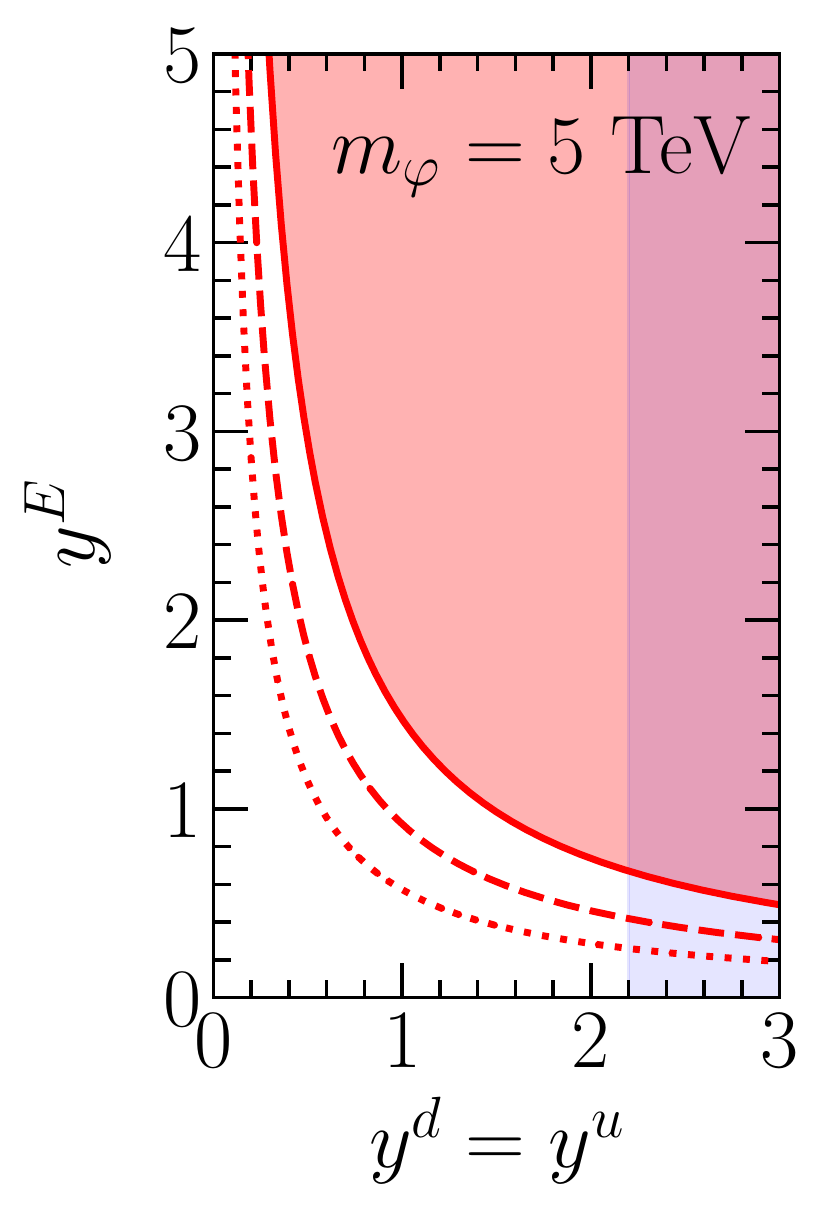}
 \includegraphics[width=0.5\columnwidth]{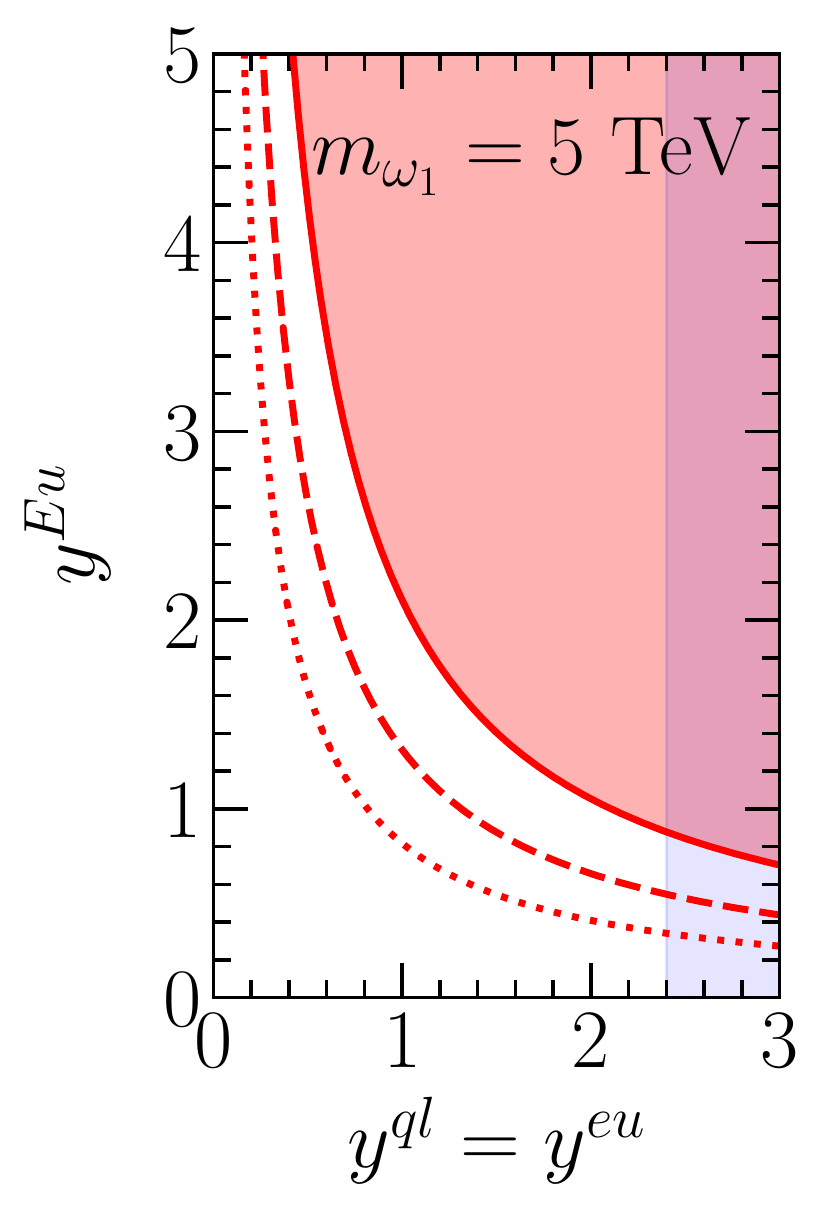}
 \includegraphics[width=0.5\columnwidth]{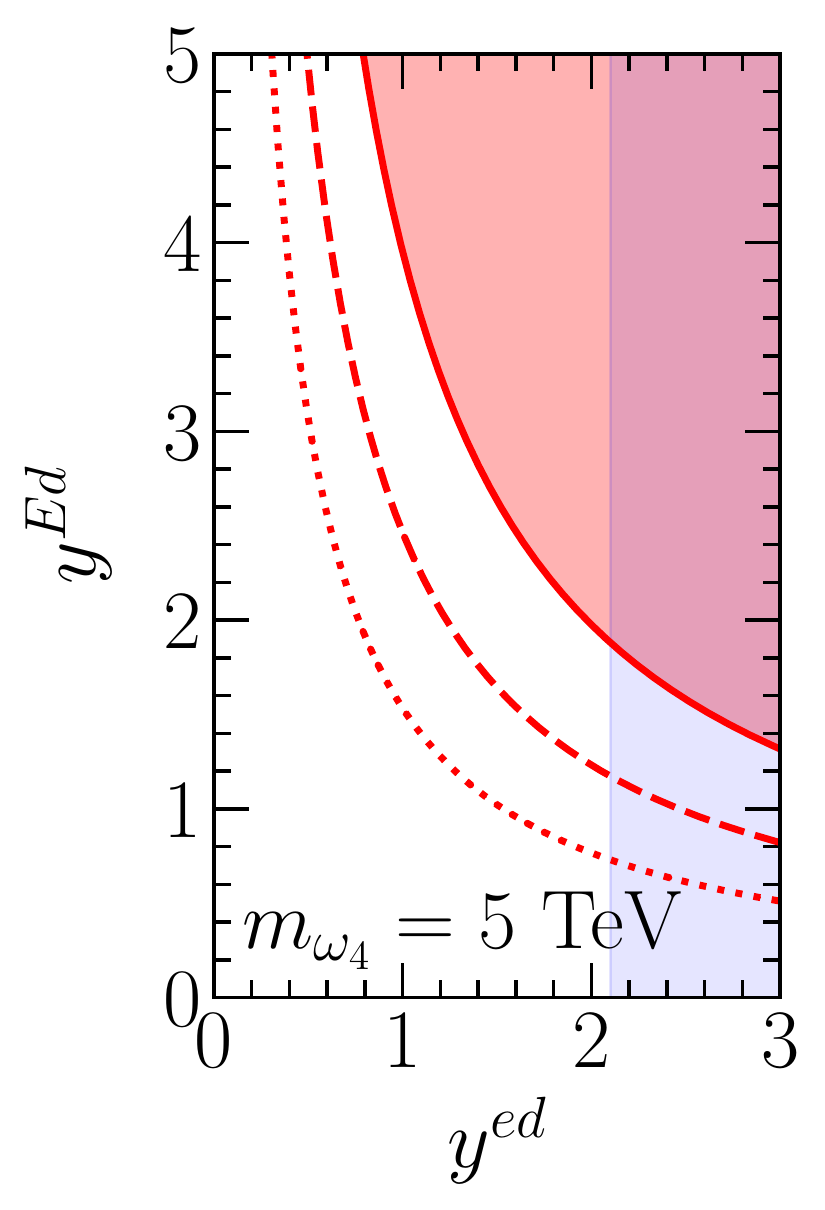}
 \includegraphics[width=0.5\columnwidth]{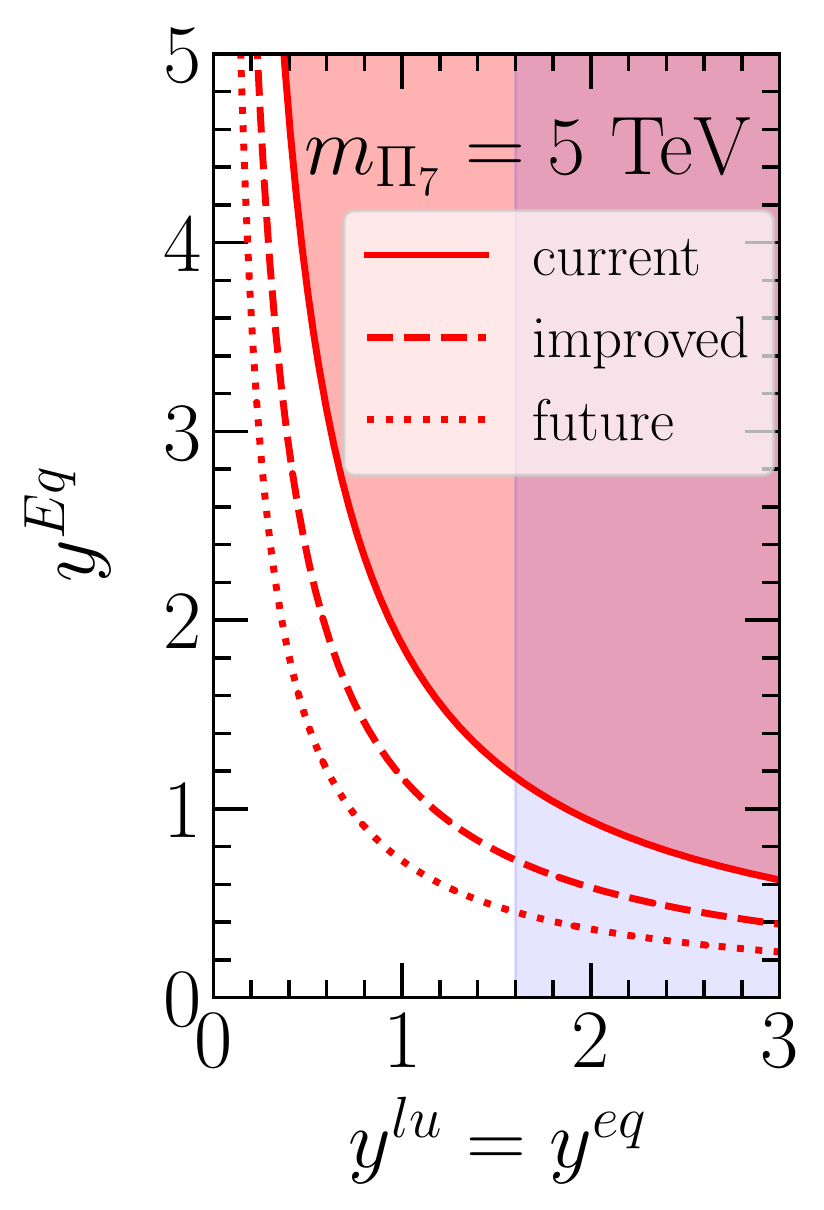}
 \caption{\it Constraints on the couplings of the scalar UV completions 
 of the ESMEFT derived under the assumption that all couplings 
 to the SM fields in a given model are equal. 
 We have used Eq.~\eqref{eq:mastereq} along with the values 
 of $\mathcal{I}$s from Tab.~\ref{tab:700GeV} and those of
 $s_\mathrm{max}$ from Tabs.~\ref{tab:500GeV} and \ref{tab:smax}, 
 assuming the light quark decay channel of $E$ and the second bin.
 ``Current'', ``improved'' (``future'') refer to the developed LHC (HL-LHC) 
 analyses described in the text. The light blue regions are excluded 
 from EWPD or dijet searches at the LHC~\cite{deBlas:2014mba}.}
 \label{fig:scalars}
\end{figure*}

Assuming for simplicity that all couplings not involving $E$ are equal, 
we show these results for the scalar mediators in Fig.~\ref{fig:scalars}. 
We also show for comparison the bounds from low-energy data 
and dijet searches~\cite{deBlas:2014mba}. 
Interestingly, \textit{e.g.} in the case of $\omega_1$, 
we see that for sufficiently large values of $y^{Eu}$, 
the bound on $y^{ql}=y^{eu}$ from our study 
is about 6 times more stringent than that from other data,  
and it can be improved by a factor of two at the HL-LHC. 
Despite not being explicitly shown, 
results for vector boson extensions of the SM$+E$ are similar. 
Let us also note 
that, even for these masses, resonant searches are not necessarily more 
constraining. The reasons are: \textit{(i)}~Several of the mediators above 
proceed 
in $t$-channel, therefore not manifesting as peaks in the distribution of the 
total invariant mass. \textit{(ii)}~Even $s$-channel mediators can in general 
decay into 
other final states, probably harder to detect (\textit{e.g.} invisible), which 
can even dominate the decay width; the EFT approach is insensitive to these 
effects. Evidently, for mediator masses above $10$ TeV, the EFT approach is 
indisputable.

In good approximation, our results can also be easily extended to four-fermion 
operators involving only second and third generation quarks. For example, due 
to the PDF suppression, the cross section for single $E$ production initiated 
by bottom quarks is about two orders of magnitude smaller than that initiated 
by down quarks. Therefore, it is expected that values of the Wilson 
coefficients $f$ ten times larger can be probed at the LHC. 
Note that the EFT is still valid in this case if we still restrict to 
$m_{\ell\ell j j}< 2.5$ TeV; see appendix~\ref{app:unitarity}.

This observation can be used to explore the sensitivity to other models. For 
concreteness, following Ref.~\cite{Chala:2018igk}, 
let us consider the SM$+E$ extension with 
a full singlet vector boson $V$ with mass $m_V$ and couplings
\begin{align}
 L =  V^\mu \Big[g_{Vqq} \overline{q}^3_L\gamma_\mu q^3_L + 
 g_{VE\ell} \left(\overline{E}\gamma_\mu \mu_R+\text{h.c.}\right)\Big]+\dots~.
 \label{eq:lfu}
\end{align}
The ellipsis encode terms not relevant for us, 
such as light lepton couplings to $V$, etc.

We fix $g_{Vqq}\sim 0.05 m_V^2/\text{TeV}^2$. 
In the original reference this value is motivated by the flavour 
anomalies~\cite{Aaij:2013qta,Aaij:2014ora,Aaij:2014pli,
Aaij:2015esa,Aaij:2015oid,Wehle:2016yoi,Aaij:2017vbb}. 
We keep the strong coupling $g_{VE\ell}$ free; 
while in the original reference it is fixed to $2.5$. 
(The phenomenology studied there is not very sensitive to the value of this 
coupling.)

\begin{figure}[t]
 \includegraphics[width=\columnwidth]{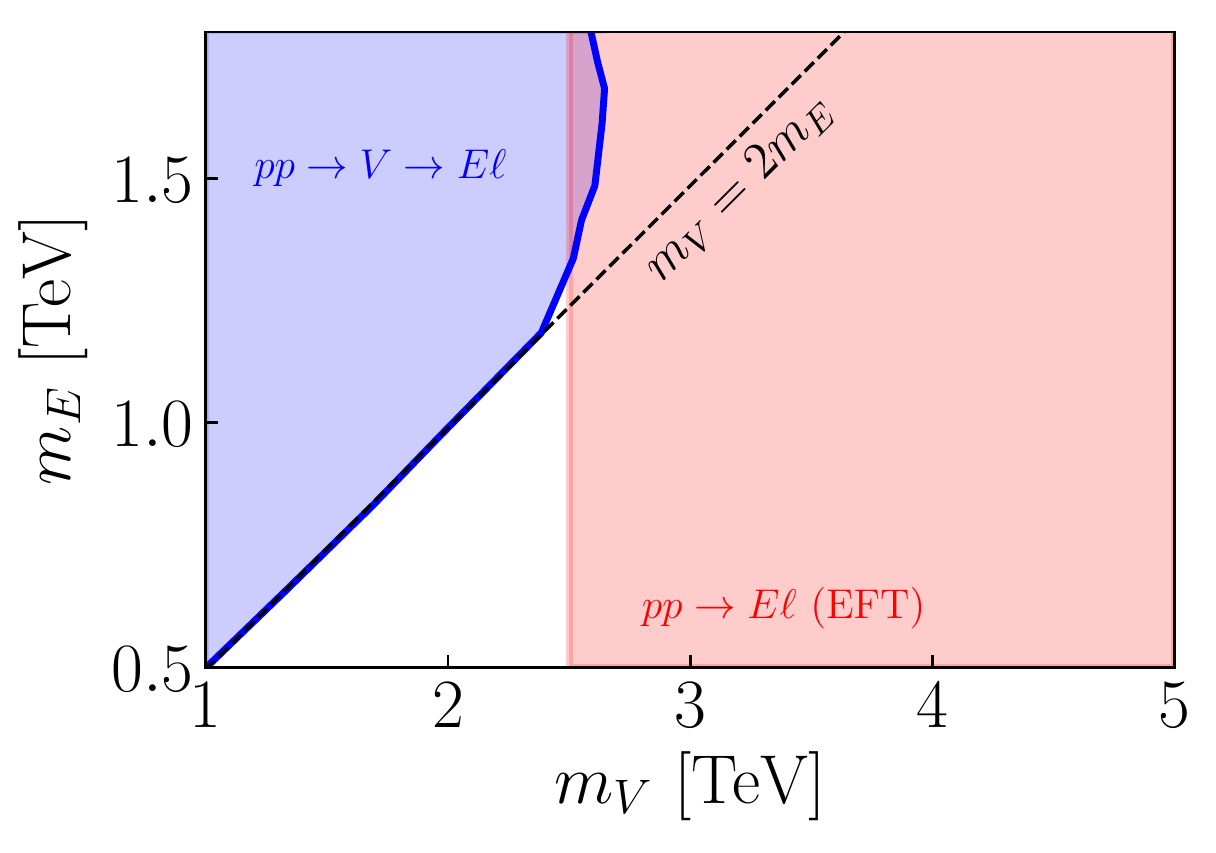}
 \caption{\it Reach of the LHC to the model described in 
 Eq.~\eqref{eq:lfu} using the resonant analysis of 
 Ref.~\cite{Chala:2018igk} (blue)
 versus the reach using the EFT analysis 
 described in this article (red); see text for details.}
 \label{fig:lfu}
\end{figure}

Upon integrating $V$ out, the only ESMEFT operator (relevant for single 
production) generated is $\mathcal{O}_{qe}$, with 
\begin{equation}
 f_{qe} = 
-\frac{g_{Vqq} g_{VE\ell}}{m_V^2}\sim -0.05 g_{VE\ell}\, \text{TeV}^{-2}\,.
\end{equation}
To compare the complementarity between our current analysis and that of 
Ref.~\cite{Chala:2018igk}, let us assume that $E$ decays equally into SM gauge 
bosons and via the four-fermion operators. Thus, the region that can be probed 
at the HL-LHC following Ref.~\cite{Chala:2018igk} (see right panel of 
Fig.~5 therein) is depicted in blue in Fig.~\ref{fig:lfu}. The area below the 
line $m_V=2m_E$, in which an on-shell produced $V$ decays into $\overline{E}E$, is not accessible within that analysis. 
Due to the resonant nature of that search, 
the region above $m_V = 2.5$ TeV remains open.

On the other hand, within our current analysis in the bin $[1.5-2.5]$ TeV, we 
can probe values of $f_{qe}$ of order $0.3$~TeV$^{-2}$ at the HL-LHC, 
which corresponds to $g_{VE\ell}\sim6$. 
(This value is significantly smaller if $E$ decays only via four-fermions; 
in which case the analysis of Ref.~\cite{Chala:2018igk} 
is not sensitive to the model.)  
Notably, this constraint is $m_V$-independent, 
provided $m_V>2.5$ TeV so that the EFT approach is valid. 
The corresponding bound is shown in red in Fig.~\ref{fig:lfu}~\footnote{We are making the conservative assumption that within our 
current analysis we are equally sensitive to values of $m_E$ above our higher 
benchmark of $m_E=900$ GeV. In light of the experimental results in 
Ref.~\cite{Sirunyan:2020awe}, it is expected that the sensitivity to 
heavier $E$ could be even
better.}.

Thus, we can conclude that for sufficiently large $g_{VE\ell}$, our analysis 
together with that in Ref.~\cite{Chala:2018igk} can completely probe the 
corresponding explanation of the flavour anomalies.

\section{Conclusions}
\label{sec:conclusions}
%
Using an effective field theory (EFT) approach, we have argued that, 
differently to what current searches for vector-like leptons (VLLs) $E$ assume, 
$E$ can be produced at high values of $\sqrt{s}$ 
via four-fermion interactions at the LHC. 
They can also decay as $E\to \ell q\overline{q}$ with \textit{no} 
intermediate Standard Model (SM) gauge bosons.

We have shown that there are other (few) experimental analyses, 
most importantly searches for excited leptons~\cite{Sirunyan:2020awe}, 
that are very sensitive to our hypothesis. 
They are however limited in scope, 
because they focus only on the case $q=u,d,c,s$ 
leaving bottom and top quarks aside, 
as well as a single four-fermion operator. 
Moreover, the statistical analysis in Ref.~\cite{Sirunyan:2020awe} 
does not apply to models with further particles below $10$ TeV.
Likewise, interpreting their bounds on ESMEFT operators 
involving only sea quarks breaks the EFT validity.
(These objections apply also to other previous similar 
analyses~\cite{Aaboud:2019zpc,Khachatryan:2015scf}.)

Thus, we have worked out the most generic base 
of EFT contact interactions involving $E$ to dimension six. 
Upon recasting the experimental analysis of Ref.~\cite{Sirunyan:2020awe},
we have obtained global bounds on all the EFT directions, for light, bottom and
top quarks separately. To this aim, we have restricted to events with $\sqrt{s}$
below the threshold determined by perturbative unitarity, that we have also
derived. Our findings show that Wilson coefficients as small 
as $0.05$~TeV$^{-2}$ are already ruled out in the muon channel.

For comparison, Ref.~\cite{Sirunyan:2020awe}, which uses all energies bins 
to $10$ TeV, reports $\Lambda \sim 20$~TeV
for couplings of order $2\pi$ (and $m_E \sim $~TeV).
This translates to $f \sim 0.015$~TeV$^{-2}$.

In the electron channel, our bounds on the Wilson coefficients 
are only slightly altered. Taking, for example, $m_E = 500$ GeV and using the 
energy bin $[1.5-2.5]$ TeV, the bounds on $f_{qe}$ for muons and for electrons read, respectively, 0.046 (0.067)~TeV$^{-2}$ and 0.048 (0.070)~TeV$^{-2}$, 
for $E$ decaying into light and bottom jets (tops).

We have also modified the current analysis with cuts on new observables 
(most importantly the number of $b$-tagged jets 
and the reconstructed mass of $E$);
improving the aforementioned bounds by a factor of 
$\sim 1.6$ ($\sim 1.4$) for light and bottom quarks (tops).

Finally, we have applied our findings to concrete UV completions 
of the SMEFT extended with $E$. 
In particular, we have classified all possible single field extensions
of the SM$+E$ that induce the four-fermion interactions of interest at tree level.
The limits on the couplings of these fields to purely SM currents can overcome 
those from low-energy data and dijet searches at the LHC by almost an order 
of magnitude with the improved analysis.

Altogether, our work motivates different searches for VLLs, 
that might be implemented by small modifications 
of current searches for excited leptons.

\section*{Acknowledgements}
%
We would like to thank Jose Santiago for useful discussions. 
MC is supported by the Spanish MINECO 
under the Juan de la Cierva programme 
as well as by the Ministry of Science and Innovation 
under grant number FPA2016-78220-C3-3-P (fondos FEDER), 
and by the Junta de Andaluc{\'i}a grant FQM 101.
PK is supported by the Spanish MINECO project 
FPA2016-78220-C3-1-P (fondos FEDER).
MC and PK also acknowledge support 
by the Junta de Andaluc{\'i}a grant A-FQM-211-UGR18 (fondos FEDER). 
MR is supported by Funda\c{c}\~ao para a Ci\^encia e Tecnologia (FCT) 
under the grant PD/BD/142773/2018 and LIP 
(FCT, COMPETE2020-Portugal2020, FEDER, POCI-01-0145-FEDER-007334).
MR also acknowledges support by FCT under project CERN/FIS-PAR/0024/2019.

\appendix
%
%
\section{Perturbative unitarity bounds}
\label{app:unitarity}
%
The aim of this appendix is to discuss the validity of the EFT approach. 
To this end, we first sum up the perturbative unitarity condition
and apply it to the tree-level EFT amplitudes $q\overline{q}\rightarrow E 
\overline{\ell}$. 
More specifically, we derive constraints on the maximum partonic
centre-of-mass energy $\hat{s}$ at which the EFT is applicable 
as a function of the Wilson coefficient $f$ of each operator. 

The unitarity of the $S$ matrix, $SS^\dagger=1$, together with	 the 
requirement of perturbativity imply that the partial waves $\mathcal{T}_j$ in 
the following partial wave decomposition of inelastic scattering amplitudes
\begin{equation}
 \mathcal{M}=16\pi\sum_J^\infty 
 (2J+1)\mathcal{T}^{(J)}_{\lambda_1',\lambda_2';\lambda_1,\lambda_2}(\hat{s}) \,d^
 {(J)\ast}_{\mu, \nu}(\theta),
 \label{eq:uni12}
\end{equation}
should fulfil the condition
\begin{equation}
 \left|\mathcal{T}^{(J)}_{\lambda_1'\lambda_2';\lambda_1\lambda_2}(\hat{s}
 )\right|\leq\frac{1}{2}
 \label{eq:uni28}
\end{equation}
for each $J\in\{J_{\mathrm{min}},J_{\mathrm{min}}+1,\ldots\}$. In this 
expression, $\lambda_{1,2}$ and  $\lambda_{1,2}'$ are the helicities of 
the initial and final particles, 
respectively; $\mu=\lambda_1-\lambda_2$, $\nu=\lambda_1' - \lambda_2'$; 
$d^{(J)}$ are the Wigner matrices in the limit of azimuthal scattering 
angle $\phi\rightarrow0$, and 
$J_{\mathrm{min}}=\mathrm{max\{|\lambda_{1}-\lambda_{2}|,|\lambda_{1}
'-\lambda_{2}'|\}}$.
For more details on the partial wave unitarity 
condition, see \textit{e.g.} Ref.~\cite{Kozow:2019ihv}.

Since the EFT amplitudes grow with the energy $\hat{s}$,
so do the partial waves. We define the distinguished energy scale 
$\sqrt{\hat{s}^U}$ as the one
 that saturates the condition in Eq.~\eqref{eq:uni28}: 
\begin{equation}
 \left|\mathcal{T}^{(J)}_{\lambda_1'\lambda_2';\lambda_1\lambda_2}(\hat{s}^U
 )\right|=\frac{1}{2}\,.
 \label{eq:uni29}
\end{equation}
Importantly, $\hat{s}^U$ is a function of $f$. Given that for energies above 
$\hat{s}^U$ the EFT amplitudes are ill-defined,  $\hat{s}^U$ 
defines the upper bound on $\hat{s}$ for which the EFT 
approach is valid.

Typically, the first partial wave yields the strongest unitarity bounds on 
$\hat{s}$. Correspondingly, we derive the bounds using 
$\mathcal{T}^{(J_{\mathrm{min}})}$. 
Since in our study the global bounds 
on $f$ are expressed in terms of each Wilson coefficient $f$ separately, 
we compute  the unitarity bounds using one operator at a time. 
The $J$-th partial wave projections are computed using 
the orthogonality of the Wigner functions:
\begin{equation}
 \mathcal{T}^{(J)}_{\mu,\nu} = \frac{1}{32\pi}\int_{-1}^1{d\cos\theta\, 
 d^{(J)}_{\mu, \nu}(\theta) }\,\mathcal{M}\,.
 \label{eq:proj}
\end{equation}

More specifically, for each operator we consider all helicity
$q\overline{q}\rightarrow E\overline{\ell}$  
amplitudes, where $q=u,d$, that are 
non-vanishing in the relativistic limit. For each such helicity combination we 
project the amplitude $\mathcal{M}$ onto the $J_{\mathrm{min}}$ partial 
wave and derive the corresponding bound on $\hat{s}$. Finally, we identify 
$\hat{s}^U$ as the lowest among all such bounds. 

Unitarity bounds for different values of $f$ are presented in Tab.~\ref{tab:pk1}. 
For example, for $f=1$ TeV$^{-2}$ the bounds are in the range 
$\sqrt{\hat{s}^U}\in[5-7]$~TeV, depending on the operator involved. 
\begin{table}[t]
 \centering
 \begin{tabular}{|c|cccccc|}
  \hline 
  $f~\left[\text{TeV}^{-2}\right]$ & $f_{ue}$& $f_{de}$& $f_{qe}$& $f_{qdl}$& $f_{qul}$ & 
  $f_{luq}$ \rule[-1.5ex]{0pt}{4ex}\\
  \hline
  \hline
  10 & 1.9 & 1.9 & 1.9 & 1.6 & 1.6 & 2.2  \\
  1 & 6.1 & 6.1 & 6.1 & 5.0 & 5.0 & 7.1  \\
  0.1 & 19 & 19 & 19 & 16 & 16 & 22 \\
  0.01 & 61 & 61 & 61 & 50 & 50 & 71 \\
  \hline
 \end{tabular}
 \caption{\it Solutions for $\sqrt{\hat{s}^U}$ 
 (in TeV) from tree-level partial wave unitarity in the presence of a 
 single operator at a time, for different values of the Wilson coefficients $f$.  
 The examined processes are 
 $u\overline{u}\to E \overline{\ell}$, 
 $d\overline{d}\to E \overline{\ell}$.}
 \label{tab:pk1}
\end{table}

For completeness, let us comment on the 
values of $c$ in $f=c/\Lambda^2$ setting 
$\Lambda=\sqrt{\hat{s}^U}$, for different values of $f$ and for each 
effective operator. Independently of the value of $f$, we obtain that 
$\sqrt{c}=6.1$  for 
$f =f_{ue}, f_{de}, f_{qe}$; $\sqrt{c} = 5$ for $f=f_{qdl}, f_{qul}$; 
and $\sqrt{c} = 7.1$ for $f=f_{luq}$.

Note that, for a fixed $f$, $\Lambda=\sqrt{\hat{s}^U}$ can be (roughly) 
identified with the upper bound on the scale $\Lambda$. 
(The new physics scale in a UV completion 
should not be significantly separated from $\sqrt{\hat{s}^U}$ 
because it is responsible for unitarization of the complete amplitudes.) 
Therefore for a given $f$, the value of $c$, 
assuming $\Lambda=\sqrt{\hat{s}^U}$, is 
an approximate upper bound on the corresponding UV coupling.

Interestingly, the aforementioned values of $c$ are \textit{(i)}~independent 
of the value of $f$ and \textit{(ii)}~in the range between $1$ and 
$4\pi$, hence indeed close to the perturbative regime (as required by the 
perturbative unitarity condition). 

Given this, a discussion on the EFT consistency of the analyses in 
sections~\ref{sec:analysis} and \ref{sec:improvements} is in order. We note 
that the larger the value of 
$f$, the stronger the unitarity bounds. Thus, in particular, 
for the largest $f$ within the limits, the $\sqrt{\hat{s}^U}$ should not be 
lower than the chosen cut-off on the (proxy) variable 
$m_{\ell\ell jj}$. Otherwise one uses events outside the validity of the EFT 
amplitudes while setting limits on the effective coefficients $f$; 
turning them to be not suitable for EFT interpretation.

In Tab.~\ref{tab:pk2} we present the unitarity bounds as function of $f$ 
for the values relevant for the $2.5$ TeV cut-off case. Comparing the 
table with Fig.~\ref{fig:globLim} one can see that all 
limits on $f$ correspond to unitarity bounds that are not lower than the 2.5 
TeV cut-off. Hence the limits are EFT interpretable. More explicitly, unitarity 
bounds $\sqrt{\hat{s}^U}$ that  \textit{e.g.} correspond to $f$ from the 
first column in Tab.~\ref{tab:improvedbounds} read
25, 22, 27, 16, 13 and 16~TeV 
for $f_{ue}$, $f_{de}$, $f_{qe}$, $f_{qdl}$, $f_{qul}$ and $f_{luq}$,
respectively. 
\begin{table}[t]
 \centering
 \begin{tabular}{|c|cccccc|}
  \hline 
  $f~\left[\text{TeV}^{-2}\right]$ & $f_{ue}$& $f_{de}$& $f_{qe}$& $f_{qdl}$& $f_{qul}$ & 
  $f_{luq}$ \rule[-1.5ex]{0pt}{4ex} \\
  \hline
  \hline
  0.3& 11 & 11 & 11 & 9.2 & 9.2 & 13 \\ 
  0.25& 12 & 12 & 12 & 10 & 10 & 14 \\
  0.2& 14 & 14 & 14 & 11 & 11 & 16 \\
  0.15& 16 & 16 & 16 & 13 & 13 & 18 \\
  0.1& 19 & 19 & 19 & 16 & 16 & 22 \\
  0.075& 22 & 22 & 22 & 18 & 18 & 26 \\
  0.06 & 25 & 25 & 25 & 20 & 20 & 29 \\ 
  0.05& 27 & 27 & 27 & 22 & 22 & 32 \\ 
  \hline 
 \end{tabular}
 \caption{\it Same as Tab.~\ref{tab:pk1} 
 but for different values of the Wilson coefficients $f$.}
 \label{tab:pk2}
\end{table}
%

\bibliography{ESMEFT_v2}

\providecommand{\href}[2]{#2}\begingroup\raggedright\begin{thebibliography}{10}

\bibitem{Aad:2015cxa}
{\scshape ATLAS} collaboration, \emph{{Search for type-III Seesaw heavy leptons
  in $pp$ collisions at $\sqrt{s}= 8$ TeV with the ATLAS Detector}},
  \href{https://doi.org/10.1103/PhysRevD.92.032001}{\emph{Phys. Rev. D}
  {\bfseries 92} (2015) 032001}
  [\href{https://arxiv.org/abs/1506.01839}{{\ttfamily 1506.01839}}].

\bibitem{Khachatryan:2015scf}
{\scshape CMS} collaboration, \emph{{Search for Excited Leptons in
  Proton-Proton Collisions at $\sqrt{s}$ = 8 TeV}},
  \href{https://doi.org/10.1007/JHEP03(2016)125}{\emph{JHEP} {\bfseries 03}
  (2016) 125} [\href{https://arxiv.org/abs/1511.01407}{{\ttfamily
  1511.01407}}].

\bibitem{Aad:2015dha}
{\scshape ATLAS} collaboration, \emph{{Search for heavy lepton resonances
  decaying to a $Z$ boson and a lepton in $pp$ collisions at $\sqrt{s}=8$ TeV
  with the ATLAS detector}},
  \href{https://doi.org/10.1007/JHEP09(2015)108}{\emph{JHEP} {\bfseries 09}
  (2015) 108} [\href{https://arxiv.org/abs/1506.01291}{{\ttfamily
  1506.01291}}].

\bibitem{Grancagnolo:2015zsh}
{\scshape ATLAS} collaboration, \emph{{Searches for leptoquarks and heavy
  leptons with the ATLAS detector at the LHC}},
  \href{https://doi.org/10.22323/1.234.0096}{\emph{PoS} {\bfseries EPS-HEP2015}
  (2015) 096}.

\bibitem{Sirunyan:2019ofn}
{\scshape CMS} collaboration, \emph{{Search for vector-like leptons in
  multilepton final states in proton-proton collisions at $\sqrt{s}$ = 13
  TeV}}, \href{https://doi.org/10.1103/PhysRevD.100.052003}{\emph{Phys. Rev. D}
  {\bfseries 100} (2019) 052003}
  [\href{https://arxiv.org/abs/1905.10853}{{\ttfamily 1905.10853}}].

\bibitem{Aad:2020fzq}
{\scshape ATLAS} collaboration, \emph{{Search for type-III seesaw heavy leptons
  in dilepton final states in $pp$ collisions at $\sqrt{s}$ = 13 TeV with the
  ATLAS detector}},  \href{https://arxiv.org/abs/2008.07949}{{\ttfamily
  2008.07949}}.

\bibitem{Sirunyan:2019bgz}
{\scshape CMS} collaboration, \emph{{Search for physics beyond the standard
  model in multilepton final states in proton-proton collisions at $\sqrt{s} =$
  13 TeV}}, \href{https://doi.org/10.1007/JHEP03(2020)051}{\emph{JHEP}
  {\bfseries 03} (2020) 051}
  [\href{https://arxiv.org/abs/1911.04968}{{\ttfamily 1911.04968}}].

\bibitem{Chala:2014mma}
M.~Chala, J.~Juknevich, G.~Perez and J.~Santiago, \emph{{The Elusive Gluon}},
  \href{https://doi.org/10.1007/JHEP01(2015)092}{\emph{JHEP} {\bfseries 01}
  (2015) 092} [\href{https://arxiv.org/abs/1411.1771}{{\ttfamily 1411.1771}}].

\bibitem{Niehoff:2015bfa}
C.~Niehoff, P.~Stangl and D.~M. Straub, \emph{{Violation of lepton flavour
  universality in composite Higgs models}},
  \href{https://doi.org/10.1016/j.physletb.2015.05.063}{\emph{Phys. Lett. B}
  {\bfseries 747} (2015) 182}
  [\href{https://arxiv.org/abs/1503.03865}{{\ttfamily 1503.03865}}].

\bibitem{Niehoff:2015iaa}
C.~Niehoff, P.~Stangl and D.~M. Straub, \emph{{Direct and indirect signals of
  natural composite Higgs models}},
  \href{https://doi.org/10.1007/JHEP01(2016)119}{\emph{JHEP} {\bfseries 01}
  (2016) 119} [\href{https://arxiv.org/abs/1508.00569}{{\ttfamily
  1508.00569}}].

\bibitem{Carmona:2015ena}
A.~Carmona and F.~Goertz, \emph{{Lepton Flavor and Nonuniversality from Minimal
  Composite Higgs Setups}},
  \href{https://doi.org/10.1103/PhysRevLett.116.251801}{\emph{Phys. Rev. Lett.}
  {\bfseries 116} (2016) 251801}
  [\href{https://arxiv.org/abs/1510.07658}{{\ttfamily 1510.07658}}].

\bibitem{Carmona:2017fsn}
A.~Carmona and F.~Goertz, \emph{{Recent $B$ physics anomalies: a first hint for
  compositeness?}},
  \href{https://doi.org/10.1140/epjc/s10052-018-6437-1}{\emph{Eur. Phys. J. C}
  {\bfseries 78} (2018) 979}
  [\href{https://arxiv.org/abs/1712.02536}{{\ttfamily 1712.02536}}].

\bibitem{Sannino:2017utc}
F.~Sannino, P.~Stangl, D.~M. Straub and A.~E. Thomsen, \emph{{Flavor Physics
  and Flavor Anomalies in Minimal Fundamental Partial Compositeness}},
  \href{https://doi.org/10.1103/PhysRevD.97.115046}{\emph{Phys. Rev. D}
  {\bfseries 97} (2018) 115046}
  [\href{https://arxiv.org/abs/1712.07646}{{\ttfamily 1712.07646}}].

\bibitem{Chala:2018igk}
M.~Chala and M.~Spannowsky, \emph{{Behavior of composite resonances breaking
  lepton flavor universality}},
  \href{https://doi.org/10.1103/PhysRevD.98.035010}{\emph{Phys. Rev.}
  {\bfseries D98} (2018) 035010}
  [\href{https://arxiv.org/abs/1803.02364}{{\ttfamily 1803.02364}}].

\bibitem{Criado:2019mvu}
J.~C. Criado and M.~Perez-Victoria, \emph{{Vector-like quarks with
  non-renormalizable interactions}},
  \href{https://doi.org/10.1007/JHEP01(2020)057}{\emph{JHEP} {\bfseries 01}
  (2020) 057} [\href{https://arxiv.org/abs/1908.08964}{{\ttfamily
  1908.08964}}].

\bibitem{Kim:2018mks}
J.~H. Kim and I.~M. Lewis, \emph{{Loop Induced Single Top Partner Production
  and Decay at the LHC}},
  \href{https://doi.org/10.1007/JHEP05(2018)095}{\emph{JHEP} {\bfseries 05}
  (2018) 095} [\href{https://arxiv.org/abs/1803.06351}{{\ttfamily
  1803.06351}}].

\bibitem{Alhazmi:2018whk}
H.~Alhazmi, J.~H. Kim, K.~Kong and I.~M. Lewis, \emph{{Shedding Light on Top
  Partner at the LHC}},
  \href{https://doi.org/10.1007/JHEP01(2019)139}{\emph{JHEP} {\bfseries 01}
  (2019) 139} [\href{https://arxiv.org/abs/1808.03649}{{\ttfamily
  1808.03649}}].

\bibitem{deBlas:2013gla}
J.~de~Blas, \emph{{Electroweak limits on physics beyond the Standard Model}},
  \href{https://doi.org/10.1051/epjconf/20136019008}{\emph{EPJ Web Conf.}
  {\bfseries 60} (2013) 19008}
  [\href{https://arxiv.org/abs/1307.6173}{{\ttfamily 1307.6173}}].

\bibitem{Redi:2013pga}
M.~Redi, \emph{{Leptons in Composite MFV}},
  \href{https://doi.org/10.1007/JHEP09(2013)060}{\emph{JHEP} {\bfseries 09}
  (2013) 060} [\href{https://arxiv.org/abs/1306.1525}{{\ttfamily 1306.1525}}].

\bibitem{Agashe:2004rs}
K.~Agashe, R.~Contino and A.~Pomarol, \emph{{The Minimal composite Higgs
  model}}, \href{https://doi.org/10.1016/j.nuclphysb.2005.04.035}{\emph{Nucl.
  Phys. B} {\bfseries 719} (2005) 165}
  [\href{https://arxiv.org/abs/hep-ph/0412089}{{\ttfamily hep-ph/0412089}}].

\bibitem{Panico:2015jxa}
G.~Panico and A.~Wulzer, \emph{{The Composite Nambu-Goldstone Higgs}},
  vol.~913. Springer, 2016,
  \href{https://doi.org/10.1007/978-3-319-22617-0}{10.1007/978-3-319-22617-0},
  [\href{https://arxiv.org/abs/1506.01961}{{\ttfamily 1506.01961}}].

\bibitem{Duerr:2013dza}
M.~Duerr, P.~Fileviez~Perez and M.~B. Wise, \emph{{Gauge Theory for Baryon and
  Lepton Numbers with Leptoquarks}},
  \href{https://doi.org/10.1103/PhysRevLett.110.231801}{\emph{Phys. Rev. Lett.}
  {\bfseries 110} (2013) 231801}
  [\href{https://arxiv.org/abs/1304.0576}{{\ttfamily 1304.0576}}].

\bibitem{Chao:2015nsm}
W.~Chao, \emph{{Symmetries behind the 750 GeV diphoton excess}},
  \href{https://doi.org/10.1103/PhysRevD.93.115013}{\emph{Phys. Rev. D}
  {\bfseries 93} (2016) 115013}
  [\href{https://arxiv.org/abs/1512.06297}{{\ttfamily 1512.06297}}].

\bibitem{Chala:2015cev}
M.~Chala, M.~Duerr, F.~Kahlhoefer and K.~Schmidt-Hoberg, \emph{{Tricking
  Landau--Yang: How to obtain the diphoton excess from a vector resonance}},
  \href{https://doi.org/10.1016/j.physletb.2016.02.006}{\emph{Phys. Lett. B}
  {\bfseries 755} (2016) 145}
  [\href{https://arxiv.org/abs/1512.06833}{{\ttfamily 1512.06833}}].

\bibitem{Sirunyan:2020awe}
{\scshape CMS} collaboration, \emph{{Search for an excited lepton that decays
  via a contact interaction to a lepton and two jets in proton-proton
  collisions at $\sqrt{s} =$ 13 TeV}},
  \href{https://doi.org/10.1007/JHEP05(2020)052}{\emph{JHEP} {\bfseries 05}
  (2020) 052} [\href{https://arxiv.org/abs/2001.04521}{{\ttfamily
  2001.04521}}].

\bibitem{deBlas:2013qqa}
J.~de~Blas, M.~Chala and J.~Santiago, \emph{{Global Constraints on Lepton-Quark
  Contact Interactions}},
  \href{https://doi.org/10.1103/PhysRevD.88.095011}{\emph{Phys. Rev. D}
  {\bfseries 88} (2013) 095011}
  [\href{https://arxiv.org/abs/1307.5068}{{\ttfamily 1307.5068}}].

\bibitem{Falkowski:2017pss}
A.~Falkowski, M.~Gonz{\'a}lez-Alonso and K.~Mimouni, \emph{{Compilation of
  low-energy constraints on 4-fermion operators in the SMEFT}},
  \href{https://doi.org/10.1007/JHEP08(2017)123}{\emph{JHEP} {\bfseries 08}
  (2017) 123} [\href{https://arxiv.org/abs/1706.03783}{{\ttfamily
  1706.03783}}].

\bibitem{Greljo:2017vvb}
A.~Greljo and D.~Marzocca, \emph{{High-$p_T$ dilepton tails and flavor
  physics}}, \href{https://doi.org/10.1140/epjc/s10052-017-5119-8}{\emph{Eur.
  Phys. J. C} {\bfseries 77} (2017) 548}
  [\href{https://arxiv.org/abs/1704.09015}{{\ttfamily 1704.09015}}].

\bibitem{Domenech:2012ai}
O.~Domenech, A.~Pomarol and J.~Serra, \emph{{Probing the SM with Dijets at the
  LHC}}, \href{https://doi.org/10.1103/PhysRevD.85.074030}{\emph{Phys. Rev. D}
  {\bfseries 85} (2012) 074030}
  [\href{https://arxiv.org/abs/1201.6510}{{\ttfamily 1201.6510}}].

\bibitem{Alwall:2014hca}
J.~Alwall, R.~Frederix, S.~Frixione, V.~Hirschi, F.~Maltoni, O.~Mattelaer
  et~al., \emph{{The automated computation of tree-level and next-to-leading
  order differential cross sections, and their matching to parton shower
  simulations}}, \href{https://doi.org/10.1007/JHEP07(2014)079}{\emph{JHEP}
  {\bfseries 07} (2014) 079} [\href{https://arxiv.org/abs/1405.0301}{{\ttfamily
  1405.0301}}].

\bibitem{Sjostrand:2014zea}
T.~Sj{\"o}strand, S.~Ask, J.~R. Christiansen, R.~Corke, N.~Desai, P.~Ilten
  et~al., \emph{{An Introduction to PYTHIA 8.2}},
  \href{https://doi.org/10.1016/j.cpc.2015.01.024}{\emph{Comput. Phys. Commun.}
  {\bfseries 191} (2015) 159}
  [\href{https://arxiv.org/abs/1410.3012}{{\ttfamily 1410.3012}}].

\bibitem{Cacciari:2011ma}
M.~Cacciari, G.~P. Salam and G.~Soyez, \emph{{FastJet User Manual}},
  \href{https://doi.org/10.1140/epjc/s10052-012-1896-2}{\emph{Eur. Phys. J.}
  {\bfseries C72} (2012) 1896}
  [\href{https://arxiv.org/abs/1111.6097}{{\ttfamily 1111.6097}}].

\bibitem{Brun:1997pa}
R.~Brun and F.~Rademakers, \emph{{ROOT: An object oriented data analysis
  framework}}, \href{https://doi.org/10.1016/S0168-9002(97)00048-X}{\emph{Nucl.
  Instrum. Meth.} {\bfseries A389} (1997) 81}.

\bibitem{Antcheva:2009zz}
I.~Antcheva et~al., \emph{{ROOT: A C++ framework for petabyte data storage,
  statistical analysis and visualization}},
  \href{https://doi.org/10.1016/j.cpc.2009.08.005}{\emph{Comput. Phys. Commun.}
  {\bfseries 180} (2009) 2499}
  [\href{https://arxiv.org/abs/1508.07749}{{\ttfamily 1508.07749}}].

\bibitem{Read:2002hq}
A.~L. Read, \emph{{Presentation of search results: The CL(s) technique}},
  \href{https://doi.org/10.1088/0954-3899/28/10/313}{\emph{J. Phys.} {\bfseries
  G28} (2002) 2693}.

\bibitem{deBlas:2014mba}
J.~de~Blas, M.~Chala, M.~Perez-Victoria and J.~Santiago, \emph{{Observable
  Effects of General New Scalar Particles}},
  \href{https://doi.org/10.1007/JHEP04(2015)078}{\emph{JHEP} {\bfseries 04}
  (2015) 078} [\href{https://arxiv.org/abs/1412.8480}{{\ttfamily 1412.8480}}].

\bibitem{delAguila:2010mx}
F.~del Aguila, J.~de~Blas and M.~Perez-Victoria, \emph{{Electroweak Limits on
  General New Vector Bosons}},
  \href{https://doi.org/10.1007/JHEP09(2010)033}{\emph{JHEP} {\bfseries 09}
  (2010) 033} [\href{https://arxiv.org/abs/1005.3998}{{\ttfamily 1005.3998}}].

\bibitem{Aaij:2013qta}
{\scshape LHCb} collaboration, \emph{{Measurement of Form-Factor-Independent
  Observables in the Decay $B^{0} \to K^{*0} \mu^+ \mu^-$}},
  \href{https://doi.org/10.1103/PhysRevLett.111.191801}{\emph{Phys. Rev. Lett.}
  {\bfseries 111} (2013) 191801}
  [\href{https://arxiv.org/abs/1308.1707}{{\ttfamily 1308.1707}}].

\bibitem{Aaij:2014ora}
{\scshape LHCb} collaboration, \emph{{Test of lepton universality using
  $B^{+}\rightarrow K^{+}\ell^{+}\ell^{-}$ decays}},
  \href{https://doi.org/10.1103/PhysRevLett.113.151601}{\emph{Phys. Rev. Lett.}
  {\bfseries 113} (2014) 151601}
  [\href{https://arxiv.org/abs/1406.6482}{{\ttfamily 1406.6482}}].

\bibitem{Aaij:2014pli}
{\scshape LHCb} collaboration, \emph{{Differential branching fractions and
  isospin asymmetries of $B \to K^{(*)} \mu^+ \mu^-$ decays}},
  \href{https://doi.org/10.1007/JHEP06(2014)133}{\emph{JHEP} {\bfseries 06}
  (2014) 133} [\href{https://arxiv.org/abs/1403.8044}{{\ttfamily 1403.8044}}].

\bibitem{Aaij:2015esa}
{\scshape LHCb} collaboration, \emph{{Angular analysis and differential
  branching fraction of the decay $B^0_s\to\phi\mu^+\mu^-$}},
  \href{https://doi.org/10.1007/JHEP09(2015)179}{\emph{JHEP} {\bfseries 09}
  (2015) 179} [\href{https://arxiv.org/abs/1506.08777}{{\ttfamily
  1506.08777}}].

\bibitem{Aaij:2015oid}
{\scshape LHCb} collaboration, \emph{{Angular analysis of the $B^{0} \to K^{*0}
  \mu^{+} \mu^{-}$ decay using 3 fb$^{-1}$ of integrated luminosity}},
  \href{https://doi.org/10.1007/JHEP02(2016)104}{\emph{JHEP} {\bfseries 02}
  (2016) 104} [\href{https://arxiv.org/abs/1512.04442}{{\ttfamily
  1512.04442}}].

\bibitem{Wehle:2016yoi}
{\scshape Belle} collaboration, \emph{{Lepton-Flavor-Dependent Angular Analysis
  of $B\to K^\ast \ell^+\ell^-$}},
  \href{https://doi.org/10.1103/PhysRevLett.118.111801}{\emph{Phys. Rev. Lett.}
  {\bfseries 118} (2017) 111801}
  [\href{https://arxiv.org/abs/1612.05014}{{\ttfamily 1612.05014}}].

\bibitem{Aaij:2017vbb}
{\scshape LHCb} collaboration, \emph{{Test of lepton universality with $B^{0}
  \rightarrow K^{*0}\ell^{+}\ell^{-}$ decays}},
  \href{https://doi.org/10.1007/JHEP08(2017)055}{\emph{JHEP} {\bfseries 08}
  (2017) 055} [\href{https://arxiv.org/abs/1705.05802}{{\ttfamily
  1705.05802}}].

\bibitem{Aaboud:2019zpc}
{\scshape ATLAS} collaboration, \emph{{Search for excited electrons singly
  produced in proton--proton collisions at $\sqrt{s}=13$ TeV with the ATLAS
  experiment at the LHC}},
  \href{https://doi.org/10.1140/epjc/s10052-019-7295-1}{\emph{Eur. Phys. J. C}
  {\bfseries 79} (2019) 803}
  [\href{https://arxiv.org/abs/1906.03204}{{\ttfamily 1906.03204}}].

\bibitem{Kozow:2019ihv}
P.~Koz\'ow, \emph{{The W and Z scattering as a probe of physics beyond the
  Standard Model: Effective Field Theory approach}}, Ph.D. thesis, Warsaw U.,
  2019.
\newblock \href{https://arxiv.org/abs/1908.07596}{{\ttfamily 1908.07596}}.

\end{thebibliography}\endgroup

\end{document}